\pdfoutput=1

\documentclass[11pt,twoside,a4paper,cmspaper,final,collab]{cms-tdr}
\def\svnVersion{379529}\def\cmsCernNoTag{CERN-EP/2016-178}\def\cmsCernDate{\today}\def\cmsMessage{Published in the Journal of High Energy Physics as \href{http://dx.doi.org/10.1007/JHEP12(2016)083}{\doi{10.1007/JHEP12(2016)083.}}}

\begin{document}\cmsNoteHeader{EXO-12-055}

\hyphenation{had-ron-i-za-tion}
\hyphenation{cal-or-i-me-ter}
\hyphenation{de-vices}
\RCS$Revision: 374321 $
\RCS$HeadURL: svn+ssh://svn.cern.ch/reps/tdr2/papers/EXO-12-055/trunk/EXO-12-055.tex $
\RCS$Id: EXO-12-055.tex 374321 2016-11-19 22:55:38Z alverson $
\newcommand{\Zmm}{\ensuremath{\Z\to\mu^{+}\mu^{-}}}
\newcommand{\Zvv}{\ensuremath{\Z\to\nu\nu}}
\newcommand{\Wlv}{\ensuremath{\PW\to \ell\nu}}
\newcommand{\Vjets}{\ensuremath{\mathrm{V}+\text{jets}}}
\newcommand{\Zjets}{\ensuremath{\Z+\text{jets}}}
\newcommand{\Wjets}{\ensuremath{\PW+\text{jets}}}
\newcommand{\Zvvjets}{\ensuremath{\Z(\nu\nu)+\text{jets}}}
\newcommand{\Wlvjets}{\ensuremath{\PW(\ell\nu)+\text{jets}}}
\newcommand{\phojets}{\ensuremath{\gamma+\text{jets}}}
\providecommand{\cmsTable}[1]{\resizebox{\textwidth}{!}{#1}}
\providecommand{\NA}{\text{---}\xspace}
\cmsNoteHeader{EXO-12-055}
\title{Search for dark matter in proton-proton collisions at 8\TeV
with missing transverse momentum and vector boson tagged jets}

\date{\today}

\abstract{
A search is presented for an excess of events with large missing transverse momentum in association with at least one highly energetic jet, in a data sample of proton-proton
collisions at a centre-of-mass energy of 8\TeV. The data correspond to an integrated luminosity of 19.7\fbinv collected by the CMS experiment at the LHC.
The results are interpreted using a set of simplified models for the production of dark matter via a scalar,
pseudoscalar, vector, or axial vector mediator.
Additional sensitivity is achieved by tagging events consistent with the jets originating from a hadronically decaying vector boson. This search uses jet substructure
techniques to identify hadronically decaying vector bosons in both Lorentz-boosted and resolved scenarios. This analysis yields improvements of 80\% in terms of excluded signal cross sections with respect to the previous CMS analysis using the same data set. No significant excess with respect to the standard model expectation is observed and limits are placed on the parameter space of the simplified models.
Mediator masses between 80 and 400\GeV in the scalar and pseudoscalar models, and up to 1.5\TeV in the vector and axial vector models, are excluded.
}

\hypersetup{%
pdfauthor={CMS Collaboration},%
pdftitle={Search for dark matter in proton-proton collisions at 8 TeV with missing transverse momentum and vector boson tagged jets},%
pdfsubject={CMS},%
pdfkeywords={CMS, physics, dark matter}}

\maketitle
\section{Introduction}

Several astrophysical observations, including those of the radial distribution of galactic rotational speeds~\cite{vandeHulst,Rubin:1980zd,Smoot:1992td} and the angular power spectrum of the cosmic microwave background~\cite{deBernardis:2000gy,Spergel:2006hy}, suggest an abundance of
a nonbaryonic form of matter in the universe. The existence of dark matter
(DM) provides some of the most compelling evidence for physics beyond the standard model (SM) of particle physics~\cite{Feng:2010gw,Porter:2011nv}.
In many theories that extend the SM, production of DM particles
is expected at the LHC. Monojet searches~\cite{Aaboud:2016tnv,monojet1,monojet2,ATLAS:2012ky,Aad:2011xw,Chatrchyan:2012me,Khachatryan:2016reg} provide
sensitivity to a wide range of models for DM production at the LHC, while mono-V
(where V=W or Z boson) searches~\cite{monolep,Aad:2014vka,Aad:2013oja,ATLAS:2014wra} target models for DM production associated with SM V-bosons. While the mono-V searches target more specific models,
they benefit from smaller SM backgrounds.
The interpretation of
results from these and other DM searches at the LHC has typically used
effective field theories that assume heavy mediators and DM production via
contact interactions~\cite{Beltran:2010ww,Goodman:2010ku,Fox:2011pm}.  The results of this analysis are
interpreted in the context of a spin-0 or spin-1 mediator decaying to a pair of DM particles,
using a set of simplified DM
models~\cite{simplified1,Buchmueller:2013dya,Buchmueller:2014yoa,Abercrombie:2015wmb} that span a broad range of mediator and DM
particle masses, for a specific benchmark point in the model parameter space.
In the limit of large mediator masses, these simplified models are well reproduced by the EFT approach.
The models provide a simplified description of DM production that is applicable across the full kinematic region accessible at the LHC.
Furthermore, within the framework of these models, a straightforward comparison can be made of the limits
obtained by LHC experiments with those of direct detection (DD) experiments.

This paper presents a search for new phenomena leading to an excess of events with least one
energetic jet and an imbalance in transverse momentum in
proton-proton collisions at a centre-of-mass energy of 8\TeV.
The data, corresponding to an integrated luminosity of 19.7\fbinv, were collected
using the CMS detector at the CERN LHC.
This was the first CMS search to target the hadronic decay modes of the
V-bosons in the mono-V channels.
The mono-V search uses techniques designed to exploit information
available in the jet's substructure when the V-boson is
highly Lorentz-boosted. Additionally, the search uses a multivariate V-tagging technique to
identify the individual jets from moderately boosted V-bosons.

The events are categorized according to the most likely origin of the jets in the
event. The signal extraction is performed by considering the missing transverse momentum
distribution in each event category, and using multiple data control regions
to constrain the dominant backgrounds. These updates to the previous CMS monojet
analysis~\cite{monojet1} yield improvements of roughly 80\% in terms of cross section
exclusion limits, using the same data set.

This paper is structured as follows: Section 2 provides a description of the
CMS detector and object reconstruction; Section 3 outlines the DM models
explored as signal hypotheses; Section 4 provides a description of
the event selection and categorization used in
the search; Section 5 describes the modelling of backgrounds used in the signal
extraction; Section 6 presents the results and interpretations in
the context of simplified models for DM production.

\section{The CMS detector and object reconstruction}

The CMS detector, described in Ref.~\cite{CMSdetector}, is a
multi-purpose apparatus designed to study high-transverse momentum ($\pt$) products of energetic
proton-proton and heavy-ion collisions.
A superconducting solenoid surrounds its
central region, providing a magnetic field of 3.8\unit{T} parallel to the beam
direction. Charged-particle trajectories are measured by the silicon pixel and
strip trackers, which cover a pseudorapidity ($\eta$) region of $\abs{\eta} < 2.5$. A
lead tungstate crystal electromagnetic calorimeter (ECAL) and a
brass and scintillator hadron calorimeter (HCAL), surround the tracking volume and
cover $\abs{\eta} < 3$. The steel and quartz-fiber Cherenkov forward
calorimeter extends the coverage to $\abs{\eta} < 5$. The CMS muon system consists
of gas-ionization detectors embedded in the steel flux-return yoke outside the
solenoid, covering $\abs{\eta} < 2.4$. The first level of the CMS trigger
system, composed of specialized hardware processors, is designed to select the most
interesting events in less than 4\mus, using information from the calorimeters
and the muon detectors. The high-level trigger processor farm is used to
reduce the recorded event rate to a few hundred events per second.

The particle-flow (PF) algorithm reconstructs and identifies each individual particle
with an optimized combination of information from the various elements of the CMS
detector~\cite{CMS-PAS-PFT-09-001,CMS-PAS-PFT-10-001}. Jets are reconstructed by the
clustering of PF objects using both the anti-\kt algorithm~\cite{Cacciari:2008gp} with 0.5 as
the distance parameter (AK5), and the Cambridge--Aachen algorithm~\cite{cajets} with
0.8 as the distance parameter (CA8). The jets used in this analysis are required to pass standard
CMS identification criteria~\cite{jec}. The jet momenta are corrected for contamination from additional
interactions in the same bunch crossing (pileup) on the basis of the observed event energy density~\cite{Cacciari:2011ma}.
Further corrections are then applied to calibrate the absolute scale of the jet energy~\cite{jec}.

The missing transverse momentum vector $\ptvecmiss$ is defined as the negative vector sum of the $\pt$
of all final state particles that are reconstructed using the PF algorithm~\cite{CMS-PAS-JME-12-002}. The magnitude of $\ptvecmiss$
is referred to as $\ETm$. Events with a large misreconstructed $\ETm$ are removed by applying quality
filters on the tracker, ECAL, HCAL, and muon detector data.

\section{Signal hypotheses}
\label{sec:signals}
The signal hypotheses  in this search are  a set of simplified
models for DM production~\cite{simplified1,Buchmueller:2013dya,Buchmueller:2014yoa}.
These models assume the existence of an additional particle, a fermionic DM
candidate, and an additional interaction that mediates the production of DM.
In particular, it is assumed that this additional interaction is mediated by a
generic spin-0 or spin-1 particle.  The interactions are characterized by four
Lagrangians, written for a Dirac-fermion DM particle $\chi$
with mass $m_{\mathrm{DM}}$, and a vector ($\PZpr$), axial vector ($\mathrm{A}$), scalar ($\mathrm{S}$), or pseudoscalar ($\mathrm{P}$) mediator with mass $m_{\mathrm{MED}}$ as,
\begin{align}
\mathcal{L}_{\text{vector}}&\supset \frac{1}{2}m_{\mathrm {MED}}^2 \PZpr_{\mu} {\PZpr}^{\mu} - g_{\mathrm {DM}}\PZpr_{\mu} \overline{\chi}\gamma^{\mu}\chi - g_{\mathrm {SM}} \sum_{\PQq} \PZpr_{\mu} \mathrm{\PAQq}\gamma^{\mu}\PQq - m_{\mathrm {DM}} \overline{\chi}\chi, \label{eq:V} \\
\mathcal{L}_{\mathrm{axial~vector}}&\supset  \frac{1}{2}m_{\mathrm {MED}}^2 \mathrm{A}_{\mu} \mathrm{A}^{\mu} - g_{\mathrm {DM}} \mathrm{A}_{\mu} \overline{\chi}\gamma^{\mu}\gamma^5\chi -g_{\mathrm {SM}}\sum_{\PQq}  \mathrm{A}_{\mu} {\PAQq}\gamma^{\mu}\gamma^{5}\PQq - m_{\mathrm {DM}} \overline{\chi}\chi, \label{eq:A} \\
\mathcal{L}_{\text{scalar}}  & \supset - \frac{1}{2} m_{\mathrm {MED}}^2 \mathrm{S}^2 - g_{\mathrm {DM}}  \mathrm{S} \overline{\chi}\chi - g_{\PQq} \sum_{{\PQq=\PQb,\PQt}} \frac{m_{\PQq}}{v} \mathrm{S} \PAQq\PQq  - m_{\mathrm {DM}} \overline{\chi}\chi,\label{eq:S} \\
\mathcal{L}_{\text{pseudoscalar}}&\supset -\frac{1}{2}m_{\mathrm {MED}}^2 \mathrm{P}^2 - ig_{\mathrm {DM}}  \mathrm{P}  \overline{\chi} \gamma^5\chi - i g_{\PQq}\sum_{{\PQq=\PQb,\PQt}} \frac{m_{\PQq}}{v}  \mathrm{P}  \mathrm{\PAQq}\gamma^{5}\PQq  - m_{\mathrm {DM}} \overline{\chi}\chi,\label{eq:PS}
\end{align}
where $v=246$\GeV is the SM Higgs potential vacuum expectation value~\cite{Harris:2014hga}.
For the vector and axial vector mediators, the terms $g_{\mathrm{DM}}$ and $g_{\mathrm{SM}}$ denote the couplings
of the mediator to the DM particle and to SM particles, respectively.
In all models considered, these couplings are assumed to be unity ($g_{\mathrm{SM}}=g_{\mathrm{DM}}=1$).
For the vector and axial vector mediators, this implies that the coupling is universal between the mediator and quarks of all flavours.
For the scalar and pseudoscalar models, $g_{\PQq}=1$ is assumed for all quark flavours, which implies
a SM Higgs-like coupling of the mediator to the SM fermions.
The split in terms of axial vector and vector mediators in the Lagrangian
parallels the existing separation in DD experiments, into spin-dependent (SD) and
spin-independent (SI) interactions; SI can refer to either vector
or scalar mediated interactions while SD interactions refer to axial vector mediated processes.
Pseudoscalar DM-nucleon interaction cross sections are suppressed at non-relativistic DM
velocities, leading to a limited sensitivity for DD experiments to this type of interaction~\cite{Haisch:2012kf,LopezHonorez:2012kv}.

\begin{figure}[htbp]
\centering
\includegraphics[width=0.49\textwidth]{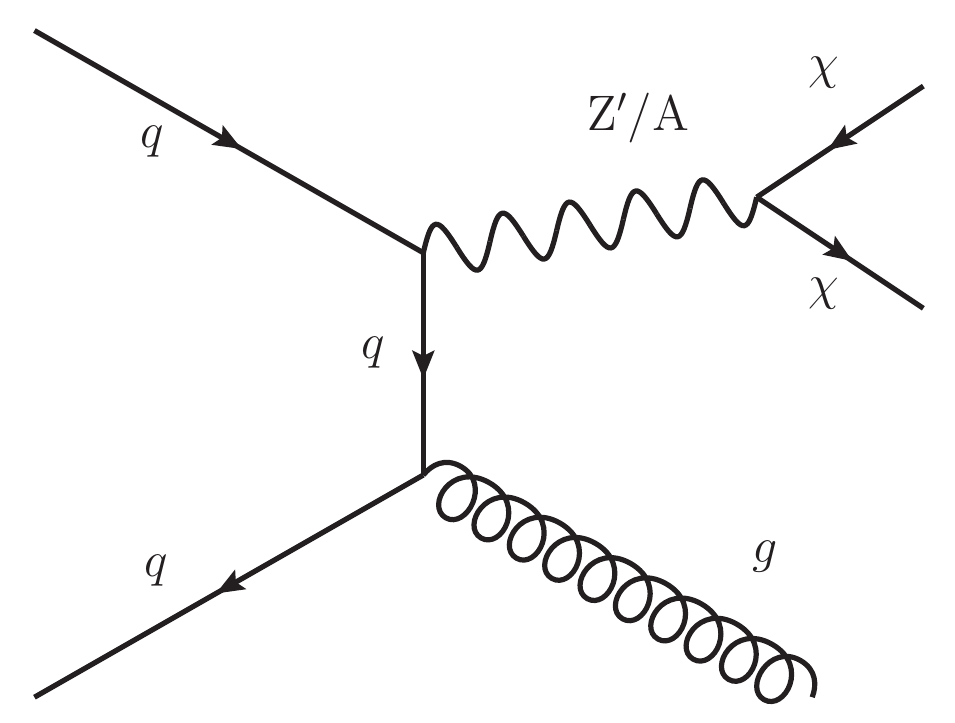}
\includegraphics[width=0.49\textwidth]{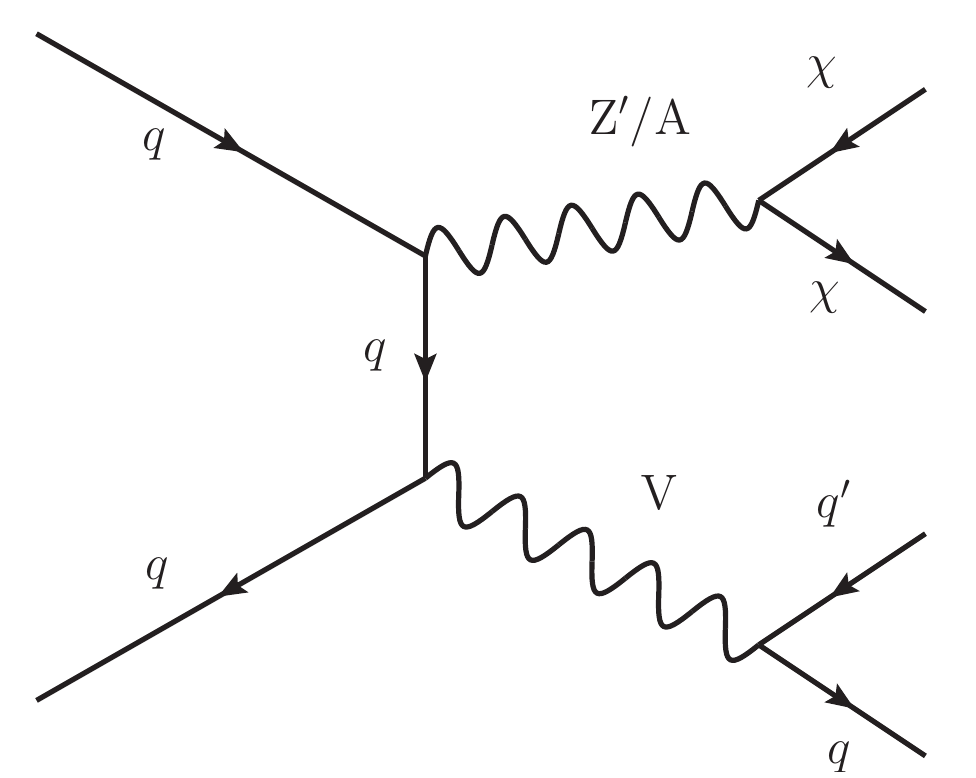}
\caption{Diagrams for production of DM via a vector ($\PZpr$) or axial vector ($\mathrm{A}$) mediator providing monojet (left) and mono-V (right) signatures.
 \label{fig:monoXfeyn1}} \end{figure}

For spin-1 signatures, the DM production process is
analogous to Z boson production via quark scattering, as shown in Fig.~\ref{fig:monoXfeyn1}. The
mono-V and monojet signatures follow from initial-state radiation (ISR) of a V-boson and
quark or gluon, respectively.
Constraints on these models for spin-1 mediators can be imposed, based on the results of searches for
visible decays of the mediator~\cite{Chala:2015ama,Alves:2013tqa,Arcadi:2013qia}, including dijet
resonance searches~\cite{ATLAS:2015nsi}. Typically, dijet resonance searches are interpreted assuming
mediator widths that are much smaller than the mediator mass~\cite{ATLAS:2015nsi,Aad:2014aqa,Khachatryan:2015sja},
while for the coupling parameter values used in this paper, the width of the spin-1 mediator is roughly 40--50\% of its mass.

\begin{figure}[htbp]
\centering
\includegraphics[width=0.49\textwidth]{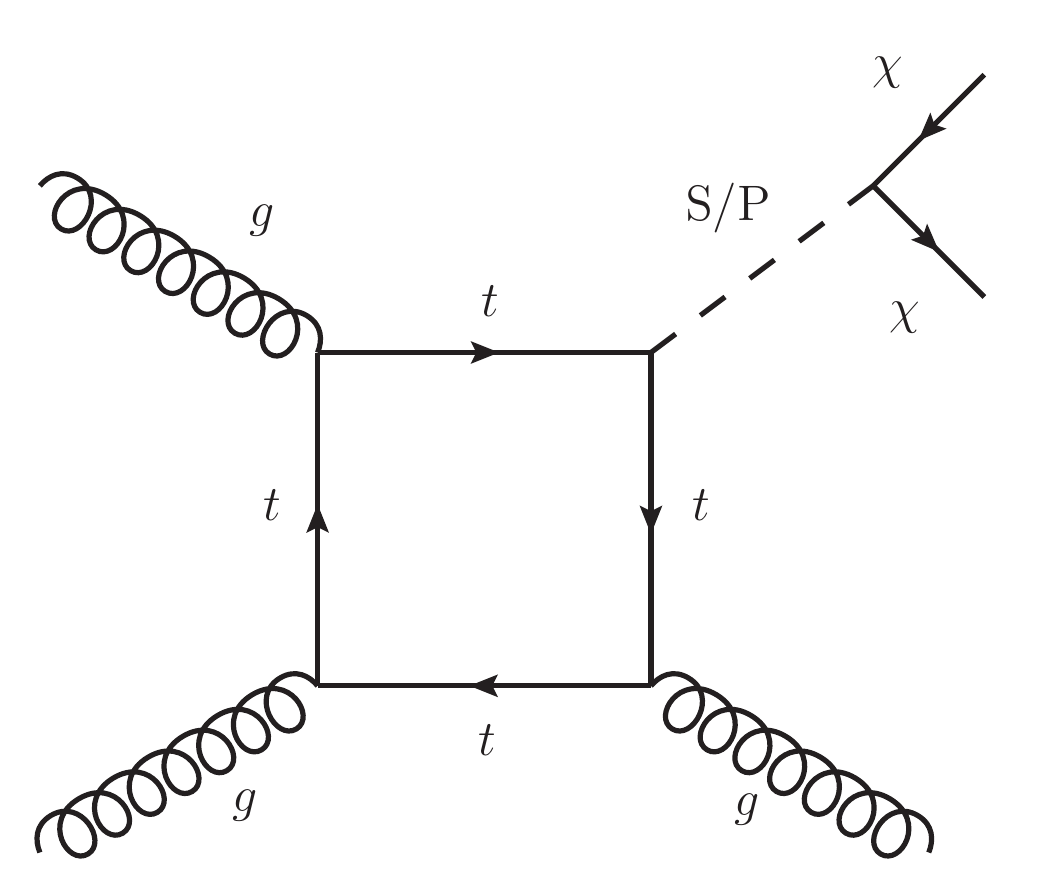}
\includegraphics[width=0.49\textwidth]{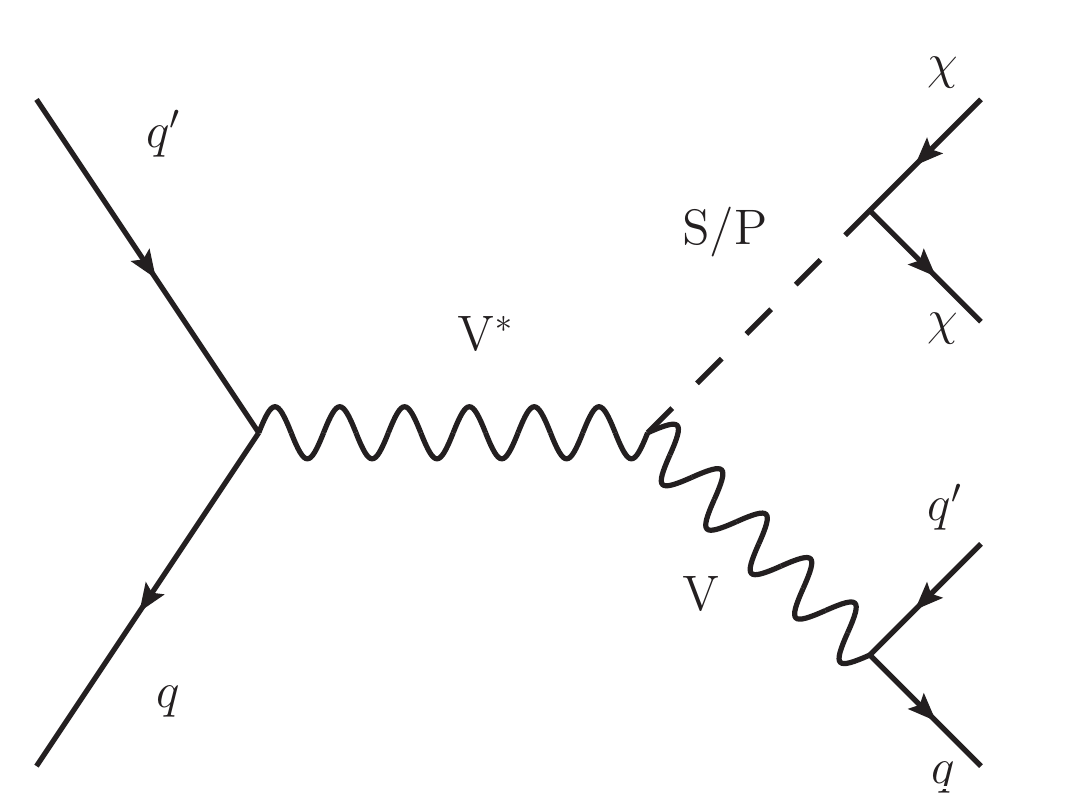}
\caption{Diagrams for production of DM via a scalar ($\mathrm{S}$) or pseudoscalar ($\mathrm{P}$) mediator in the cases
  providing monojet (left) and mono-V (right) signatures.
  \label{fig:monoXfeyn}}
  \end{figure}

The scalar and pseudoscalar models can be extended by allowing the scalar and pseudoscalar interactions to undergo
electroweak symmetry breaking in an analogous way to the Higgs
mechanism~\cite{Khoze:2015sra,Hambye:2013sna,Khoze:2014xha,Khoze:2014woa,Altmannshofer:2014vra,Carone:2013wla,Heikinheimo:2013fta}.
In such spin-0 models, the coupling of the mediator to SM quarks can be mass-dependent as parameterized in Eqs. (\ref{eq:S}) and (\ref{eq:PS}).
In these models, the production of DM at hadron colliders occurs
predominantly through gluon-fusion via a top quark loop as shown in Fig.~\ref{fig:monoXfeyn} (left).  When couplings of the mediator to vector bosons are
present, mono-V signatures are produced through a radiative process, as indicated in Fig.~\ref{fig:monoXfeyn} (right).
The scenario in which couplings between the mediator and vector bosons are not considered, is
denoted herein as \textit{fermionic}. For fermionic models, the mediator width is calculated assuming that  it couples only to quarks
and DM particles. This is referred to as the minimal width constraint. For the case in
which couplings between the mediator and V-bosons are allowed, the width is modified to
account for the additional contributions that arise~\cite{Harris:2014hga}.

To model the contributions expected from these signals, simulated
events are generated, at leading order (LO) precision, using \textsc{mcfm} 6.8~\cite{mcfm} for
the monojet signature and \textsc{JHUGen} 5.2.5 \cite{Anderson:2013afp} for the mono-V signature.
Large modifications to high-\pt production of a spin-0 mediator, produced via gluon-fusion in
association with jets, are expected when the actual mass of the top quark is used, rather than assuming
it to be infinite~\cite{Harlander:2012hf,Neumann:2014nha}.
This effect is taken into account in the generation of the scalar and pseudoscalar signals and in the calculation
of their cross sections.
The {NNPDF3.0} set of parton distribution functions (PDF) is used to specify the inputs in the signal generation~\cite{Ball:2014uwa}.
The generated events are interfaced with {\PYTHIA} 6.4.26~\cite{Sjostrand:2006za} for parton showering
and hadronization with the underlying event tune Z2*~\cite{Chatrchyan:2013ala}.
For the monojet signal, the generation is performed using the mediator
mass for the renormalization and factorization scales.
The mediator mass is also used for the scale in the parton showering (PS).

Higher-order QCD and electroweak effects are not considered in the generation of
the monojet signal. Alternative signal samples for the spin-1 mediators, generated
with {\POWHEG 2.0}~\cite{Nason:2004rx,Frixione:2007vw,powheg,Alioli:2010xd,Alioli:2009je} at next-to-leading order (NLO) precision,
followed by {\PYTHIA 8.212}~\cite{Sjostrand:2007gs} with the underlying event tune {CUETP8M1}~\cite{Khachatryan:2015pea}
for the description of fragmentation and hadronization, have been considered. The mediator \pt is used, instead of the
mediator mass, as the choice for the renormalization, factorization, and PS scales. Using the alternative samples results
in a reduction in the expected signal yield of up to 80\% for the spin-1 mediators with $m_{\mathrm{MED}} > 400$\GeV.
Signal samples for the spin-0 mediators were also generated with {\POWHEG} (at LO precision) with the
same scale choices as used for spin-1 samples. Using these samples results in a reduced signal yield
in the relevant kinematic region, by up to 30\% when $m_{\mathrm{MED}} < 400$\GeV.
The reduction in signal yields predicted by the alternative samples translates to a reduction of the exclusion
in the mediator mass by approximately 200 and 20\GeV for small $m_{\mathrm{DM}}$ values, for the spin-1 and spin-0
mediators, respectively.
Higher-order electroweak effects are expected to reduce the yield of the mono-V signal, for spin-0 mediators, by up to 15\% at large
mediator \pt~\cite{Denner:2011id}, while NLO QCD corrections are expected to increase the yield by roughly 25\%~\cite{Brein:2003wg}.

To compute the SM background expectation, simulated samples are
produced at LO for the $\Zjets$, $\Wjets$, $\ttbar$,
and QCD multijet processes using
{\MADGRAPH5.1.3}~\cite{amcatnlo} interfaced with {\PYTHIA 6.4.26} for
hadronization and fragmentation, where jets from the matrix element calculations are matched
to the parton shower, following the MLM matching
prescription~\cite{Mangano:2006rw}. Additionally, a single top quark background
sample is produced at NLO with {\POWHEG1.0},
and a set of diboson and $\phojets$ samples are
produced at LO with {\PYTHIA 6.4.26}. All of the simulated background samples are generated using
the {CT10} PDF set~\cite{Gao:2013xoa}.
The underlying event description is provided by the Z2* tune in the signal and background simulation.

The generated signal and background events are interfacted with {\GEANTfour}~\cite{geant4} to simulate the CMS detector response.
The simulated samples are then corrected to
account for the distribution of pileup interactions observed in the 8\TeV data set. All signal and background samples are
additionally corrected to account for the observed mismodelling of hadronic recoil in
simulation, following the procedure described in Ref.~\cite{CMS-PAS-JME-12-002}.

\section{Event selection and categorization}\label{sec:selection}

Candidate signal events are selected by requiring large values of
\ETm and one or more high-\pt jets.
The data used for this analysis are collected using two \ETm triggers.
The first requires $\ETm>120$\GeV, where the \ETm is calculated using a PF
reconstruction algorithm that only uses information from the calorimeters,
while the second requires $\ETm>95$\GeV or $\ETm>105$\GeV, depending on
the data taking period, together with at least one jet with $\pt>80$\GeV and
$\abs{\eta}<2.6$.

Selected events are required to have $\ETm>$200\GeV to ensure a trigger efficiency greater
than 99\% for all events used in the analysis.
The azimuthal angle $\phi$ between the $\ptvecmiss$ and the highest-\pt (leading) jet, $\abs{\Delta\phi(\ptvecmiss,\mathrm{j})}$, is
required to be larger than 2\unit{radians} to reduce the contribution from QCD multijet events.
Events are vetoed if they contain at least one well-identified electron, photon, or muon with
$\pt>10$\GeV, or a $\tau$ lepton with
$\pt>15$\GeV~\cite{Khachatryan:2015iwa,Khachatryan:2015hwa,Khachatryan:2015dfa,Chatrchyan:2013sba}. The
electron, $\tau$ lepton, and photon vetoes require that the
identified object be isolated, by using standard PF isolation algorithms~\cite{Beaudette:2014cea}.

Selected events are classified according to the
topology of the jets to distinguish between ISR of a quark or gluon,
and hadronic V-boson decays, which can be either highly Lorentz-boosted or resolved into two jets.
This approach results in three independent classes of events that are referred to as the monojet, V-boosted,
and V-resolved categories. The V-boosted and V-resolved categories are collectively referred to as the V-tagged categories.

\begin{figure}[hbtp]
\centering
{\includegraphics[width=0.49\textwidth]{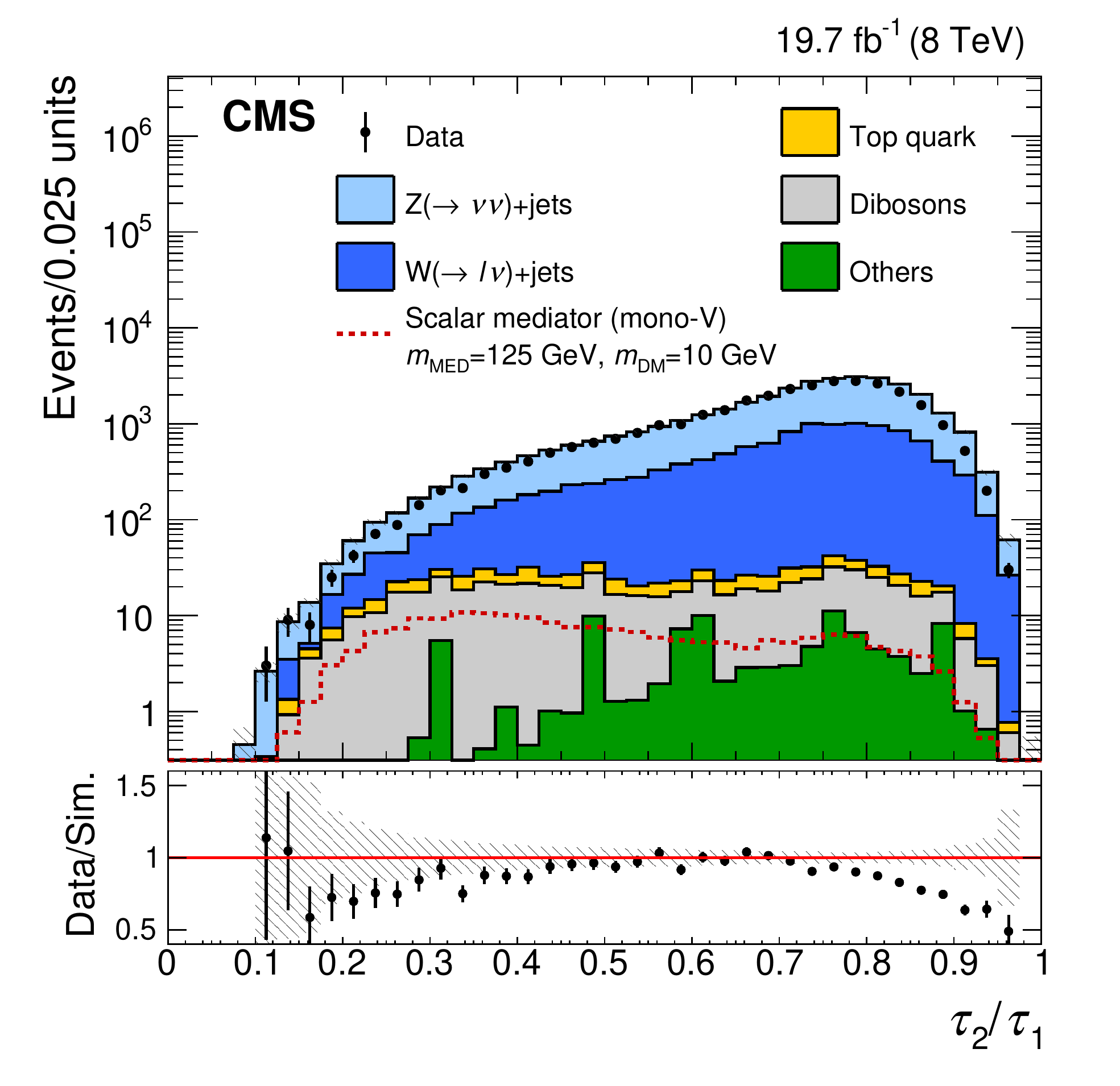}}
{\includegraphics[width=0.49\textwidth]{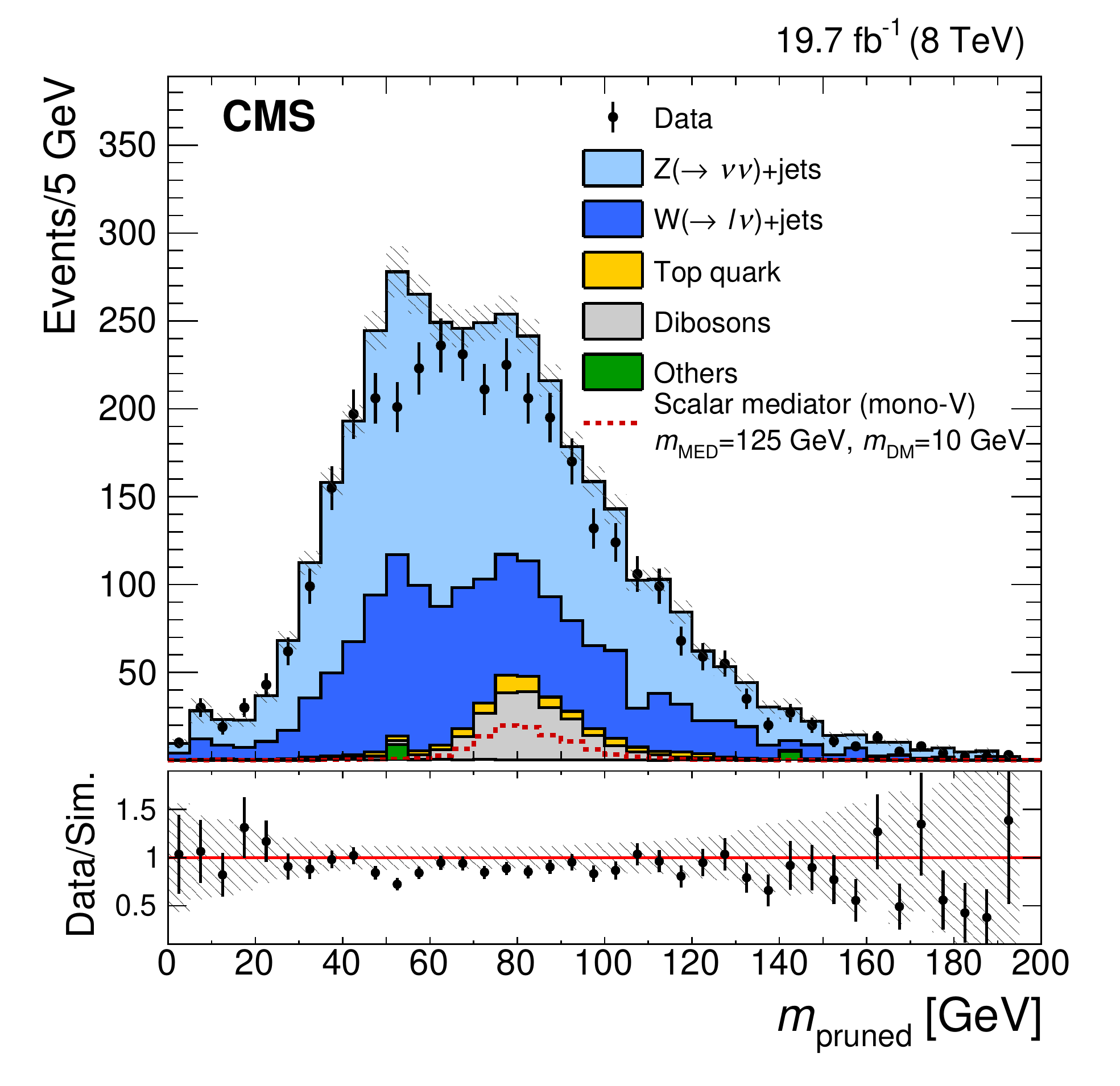}}
\caption{
Left: The distribution of $\tau_2/\tau_1$ in highly Lorentz-boosted events, before the jet mass selection.
Right: The distribution of $m_{\text{pruned}}$ for the CA8 jets, before applying the jet mass selection
but after the requirement of $\tau_2/\tau_1 < $ 0.5 has been applied.
The discrepancy between data and simulation is within systematic uncertainties (not shown).
The dashed red line shows the expected distribution for scalar-mediated DM production with
$m_{\mathrm{MED}} = 125$\GeV and $m_{\mathrm{DM}} = 10$\GeV. The shaded bands
indicate the statistical uncertainty from the limited number
of simulated events.}
\label{fig:boostvtagvars}\end{figure}

If the V-boson decays hadronically and has sufficiently large~\pt, both of its hadronic decay products are captured as
a single reconstructed ``fat'' jet.  Events in this V-boosted category are
required to have a reconstructed CA8 jet with $\pt>200$\GeV and $\ETm>250$\GeV.
Additional selection criteria are applied to improve the vector boson jet purity by
cutting on the ``N-subjettiness'' ratio $\tau_2/\tau_1$ as defined
in Refs.~\cite{Thaler:2010tr,Thaler:2011gf}, which identifies jets with a two-subjet
topology, and on the pruned jet mass ($m_{\text{pruned}}$)~\cite{Ellis:2009me}.
The $\tau_2/\tau_1$ ratio is required to be smaller than 0.5 and $m_{\text{pruned}}$
is required to be in the range 60--110\GeV.
Events which contain additional AK5 jets close to the CA8 jet, but no closer than $\Delta R=\sqrt{\smash[b]{(\delta\eta)^{2}+(\delta\phi)^{2}}} =
0.5$, are selected to include the frequent cases in which ISR
yields additional jets. If exactly one AK5 jet with $\pt>30$\GeV and $\abs{\eta}<2.5$
is reconstructed with $\Delta R> 0.5$ relative to the CA8 jet, and the azimuthal angle between it and the CA8 jet
is smaller than 2 radians, the event is selected. Events
with more than one AK5 jet with $\pt>30$\GeV and $\abs{\eta}<2.5$, reconstructed
at $\Delta R> 0.5$ relative to the CA8 jet, are rejected.
Figure~\ref{fig:boostvtagvars} shows the distributions in $\tau_2/\tau_1$ and
$m_{\mathrm{pruned}}$ before the application of the jet mass selection, in simulation
and data, for the V-boosted category. A discrepancy is present in the simulation relative to
the data. This discrepancy has been studied and found to fall within the variations observed
when using different parton shower models and detector descriptions in the simulation~\cite{Khachatryan:2014vla}.
The disagreement is within the systematic uncertainties of the
selection efficiency that are included in this analysis.

\begin{figure}[hbtp]\centering
{\includegraphics[width=0.49\textwidth]{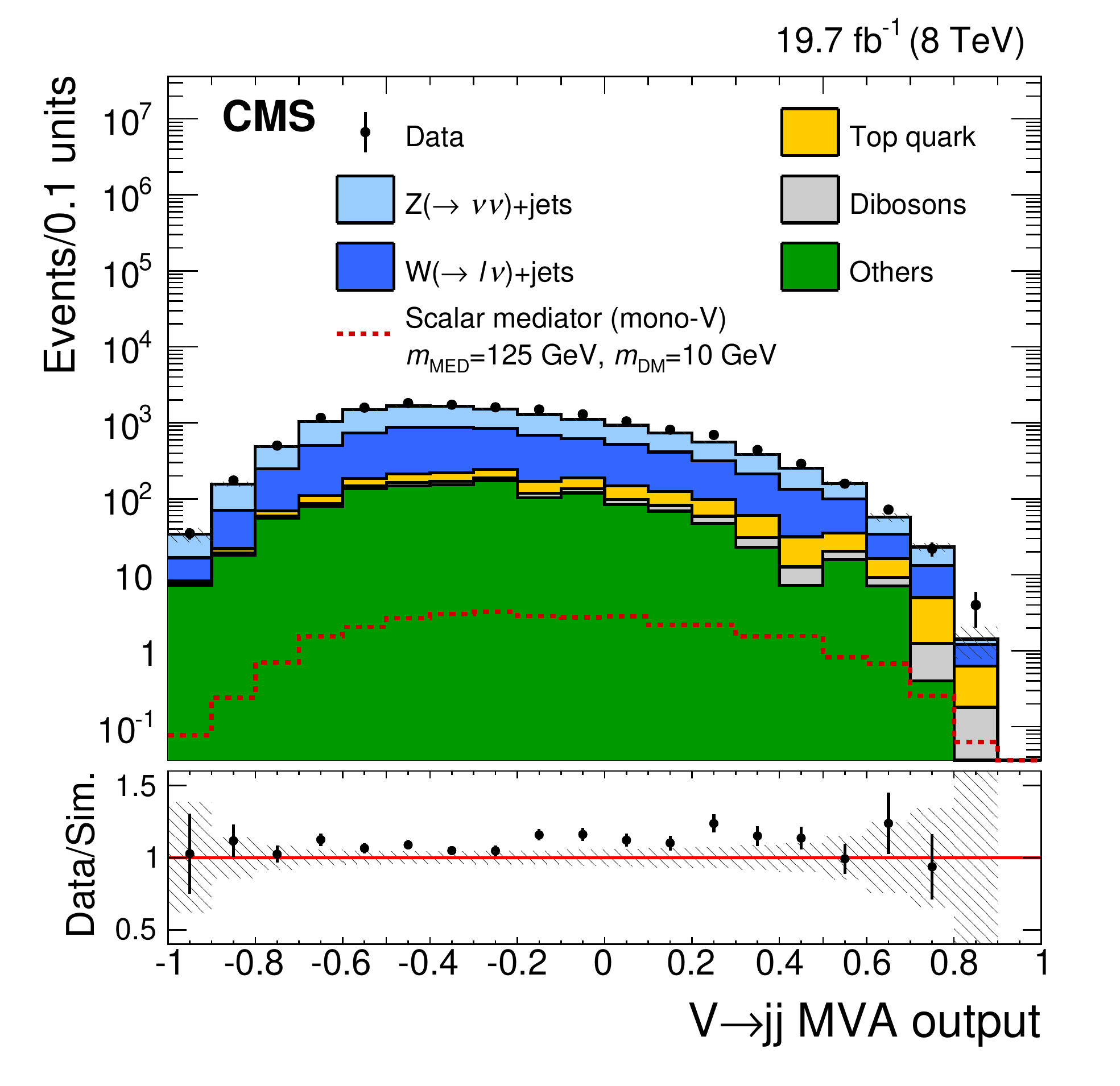}}
{\includegraphics[width=0.49\textwidth]{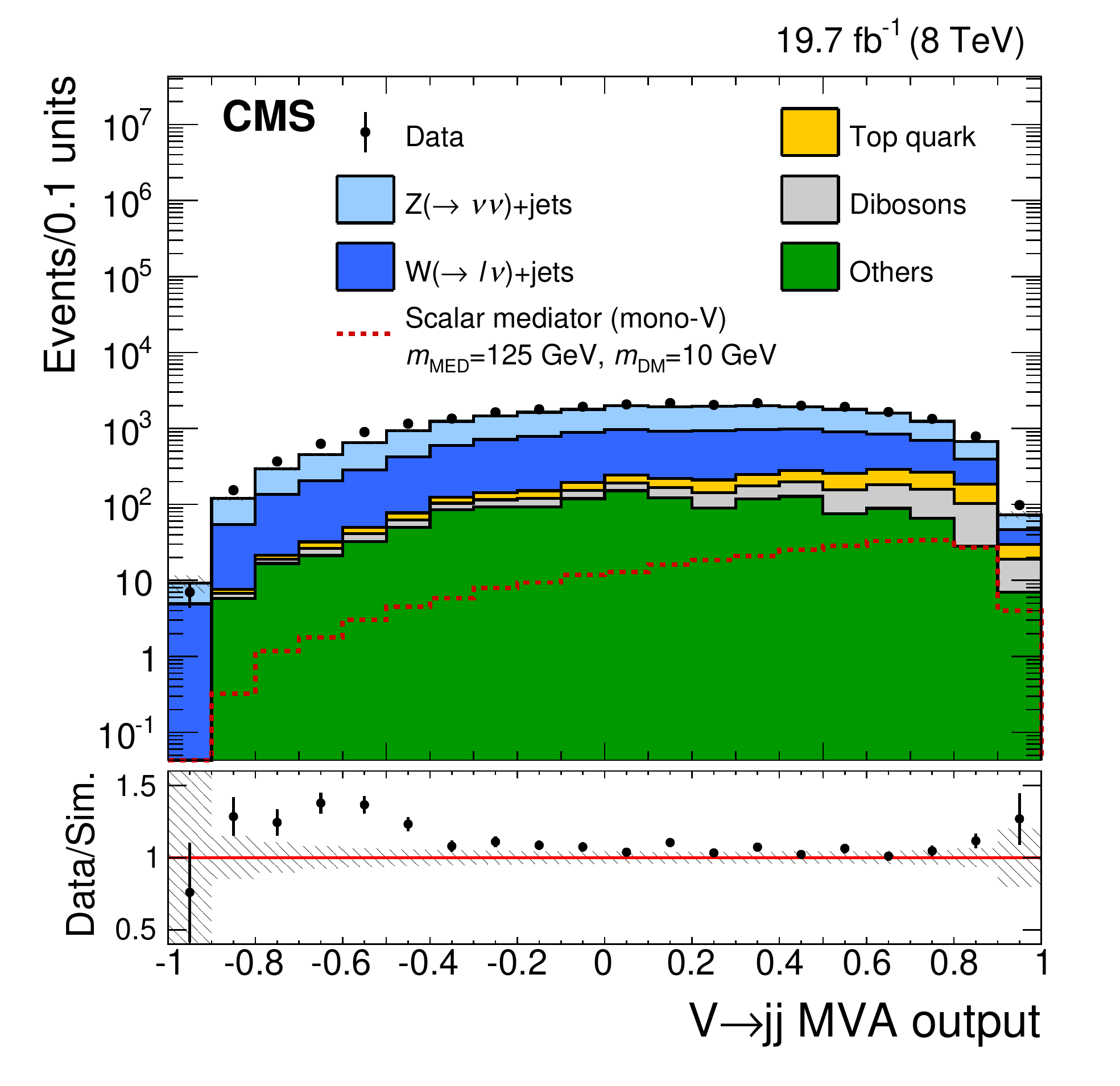}}
\caption{The MVA output distributions for V-tagged events in simulation and
data after signal selection for $\pt < 160$\GeV (left) and $\pt > 160$\GeV (right). Above a $\pt$ of about 160\GeV,
the jets from the V-boson decay begin to overlap. The dashed red line shows the expected
distribution for scalar-mediated DM production with $m_{\mathrm{MED}} = 125$\GeV and $m_{\mathrm{DM}} = 10$\GeV .
The shaded bands indicate the statistical uncertainty arising from
the limited number of simulated events.}
\label{fig:vtagger}
\end{figure}

In cases where the V-boson has insufficient boost for
its hadronic decay to be fully contained in a single reconstructed CA8 jet, a
selection that targets V-boson decays into a pair of AK5 jets is applied to recover
events failing the V-boosted selection. This selection requires that each jet
has $\pt>30$\GeV and $\abs{\eta}<2.5$, and that the dijet system has a mass in the range
60--110\GeV, consistent with originating from a W or Z boson. To reduce
the combinatorial background in this V-resolved category, a multivariate (MVA)
selection criterion is applied. The inputs to the MVA are the jet pull angle~\cite{Gallicchio:2010sw}, the mass drop variable~\cite{Izaguirre:2014ira}, and a likelihood-based discriminator that
distinguishes quark-originated from gluon-originated jets~\cite{JME-14-002}.
In events where multiple dijet pairs are
found, the pair with the highest MVA output value is taken as the candidate. The distributions of the MVA output for SM backgrounds and for a scalar
mediator produced in association with a V-boson are shown in
Fig.~\ref{fig:vtagger}. The disagreement observed between the data and simulation is included as a
systematic uncertainty in the efficiency of the V-resolved category selection for the top quark and diboson backgrounds.
Events are included in the V-resolved category if they have an
MVA output greater than 0.6. This selection is optimal for mono-V signals with a spin-0 mediator with $m_{\mathrm{MED}}<300$\GeV~\cite{JME-14-002}.

To reduce contamination from top quark backgrounds, events are rejected if they
contain a jet that is identified as a b jet, defined using the combined secondary vertex tagger
operating at a medium efficiency working point~\cite{BTAG}. Finally, the events are required to have $\ETm>250$\GeV.

Events that do not qualify for either of the two V-tagged categories are
required to have one or two high-\pt jets that are consistent with originating from a single quark or gluon. This final category is referred to
as the monojet category. For the monojet category, events are required to have $\ETm>200$\GeV and contain at
least one AK5 jet within $\abs{\eta}<2$ with $\pt >150$\GeV.
Events containing a second AK5 jet with
$\pt>30$\GeV and $\abs{\eta}<2.5$ are selected, providing the azimuthal angle between the leading jet with $\abs{\eta}<2$ and this second AK5 jet is less than 2 radians.
This selection recovers the frequent cases where
ISR yields two jets in the monojet signal.  Events with three or more AK5 jets
with $\pt>30$\GeV and $\abs{\eta}<2.5$ are rejected. Table~\ref{tab:selection} gives a
summary of the event selection in the three categories. The priority for event selection
is that events are first selected in the V-boosted category, followed by the
V-resolved category, and finally in the monojet category.
Events which pass a given selection are not included in any subsequent category.

\begin{table}[hbt]
\centering
	\topcaption{Event selections for the V-boosted, V-resolved, and monojet categories
	The requirements
	on $\pt^{\mathrm{j}}$ and $\abs{\eta}^{\mathrm{j}}$ refer to the highest $\pt$ CA8 or AK5 jet in the
        V-boosted or monojet categories, and to both leading AK5 jets in the V-resolved category. The requirement on the number of
	jets ($N_\mathrm{j}$) is applied in the V-boosted and monojet categories. An additional jet is allowed only if it falls within $\abs{\Delta\phi}<2$ radians of the
leading AK5 or CA8 jet for the monojet or V-boosted category. The additional AK5 jets in the V-boosted category must be further than $\Delta R>0.5$ for the event to fail this criteria.
	}
 \label{tab:selection}
\begin{tabular}{l|c c c}
 				     & V-boosted  & V-resolved & Monojet   \\
 \hline
  $\pt^{\mathrm{j}}$     	     & $>$200\GeV & $>$30\GeV  & $>$200\GeV  \\
  $\abs{\eta}^{\mathrm{j}}$     	     & $<$2.5     & $<$2       & $<$2  \\
  $\ETm$     		     	     & $>$250\GeV & $>$250\GeV & $>$200\GeV  \\
  $\tau_2/\tau_1$      		     &  $<$0.5          &\NA          &\NA\\
  $\mathrm{V}\to{\mathrm{jj}}$ MVA output    & \NA              & $>$0.6      &\NA\\
  $m_{\mathrm{pruned}}$    	     & 60--110\GeV &  \NA &\NA \\
  $m_{\mathrm{jj}}$    		     &\NA& 60--110\GeV &\NA \\
  $|\Delta\phi(\ptvecmiss,\mathrm{j})|$      	     & $>$2\unit{rad}     &\NA & $>$2\unit{rad} \\
  $\mathrm{N}_{\mathrm{j}}$          & $=$1     &\NA & $=$1 \\
 \hline
 \end{tabular}
\end{table}

\begin{figure}[hbtp]
\centering
{\includegraphics[width=0.49\textwidth]{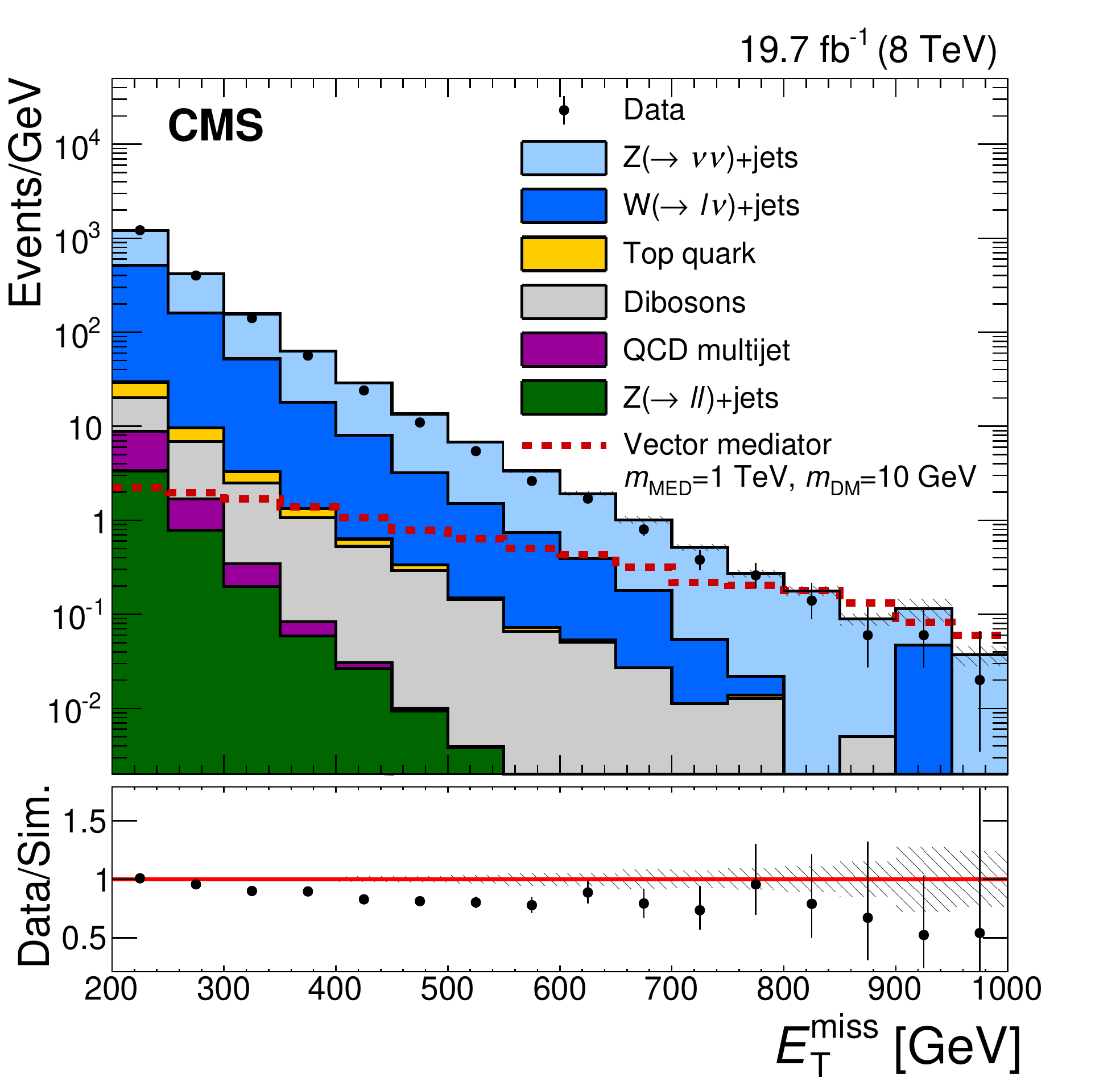}}
{\includegraphics[width=0.49\textwidth]{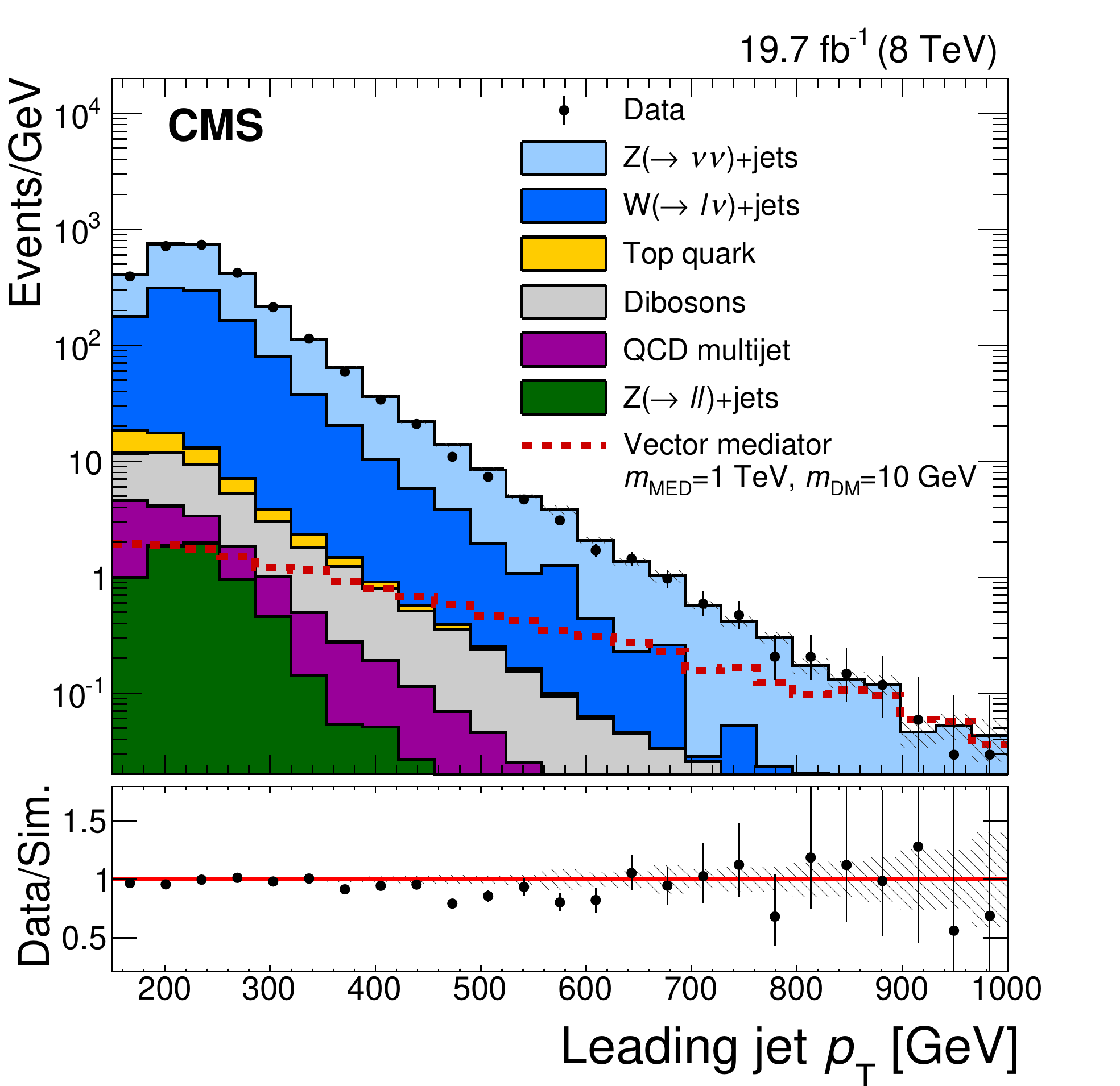}}
\caption{ Distributions in \ETm (left) and leading jet $\pt$ (right) in simulated
events and data, resulting from the combined signal selections for the three event categories.
The dashed red line shows the expected distribution, assuming vector
mediated DM production with $m_{\mathrm{MED}}=1$\TeV and $m_{\mathrm{DM}}=10$\GeV.
The shaded bands indicate the statistical uncertainty from the limited number
of simulated events.} \label{fig:ptandmet}\end{figure}

Figure~\ref{fig:ptandmet} shows the \ETm and leading jet $\pt$ distributions
in data and simulation after selection for the three event classes combined. The
backgrounds are normalized to the integrated luminosity of the data samples,
and the expected distribution for
vector mediated DM production assuming $m_{\mathrm{DM}}=$10\GeV and $m_{\mathrm{MED}}=$1\TeV
is overlaid.
The discrepancy between the data and simulation is a result of
both detector resolution and an imperfect theoretical
description of the kinematics of the $\Vjets$ processes. Both effects are
corrected using control samples in data, as described in the
following section.

\section{Background estimation}

The presence of DM production would be observable as an excess of events above SM
backgrounds at high $\ETm$. The sensitivity obtained by considering
the shape of the $\ETm$ spectrum in these events is significantly better than that achieved in the simple
counting analysis described in the previous CMS paper~\cite{monojet1}.
Additional improvement is achieved by using control regions in data to
reduce the uncertainties in the predictions of the SM backgrounds. These regions are
statistically independent from the signal region and designed such that the expected contribution
from a potential signal is negligible.
A binned likelihood fit is performed in the ranges 250--1000\GeV and 200--1000\GeV
for the two V-tagged and monojet categories, respectively. The binning is chosen to
ensure that each corresponding bin of a set of control regions is populated. The
width of the highest $\ETm$ bin is chosen to provide ease of comparison with the previous
CMS search~\cite{monojet1}.

The background contributions from $\Zvvjets$ and $\Wlvjets$ are determined using data
from dimuon and photon, and single muon control regions, respectively. The events in the control regions are
divided into the three categories, using the selection criteria
described in Section~\ref{sec:selection}, but replacing the lepton and photon vetoes with a
requirement of the presence of one of the following:
a pair of oppositely charged muons consistent with a Z boson decay, a high
$\pt$ photon, or a single muon consistent with a leptonic W boson decay. This
yields a total of nine control regions; three for each event category.
In the control regions, the transverse momentum of the dimuon pair, the
single muon, or the photon is removed and the \ETm~is recalculated. This quantity is
referred to as pseudo-\ETm~and it is this variable to which the $\ETm$ selection of the
corresponding signal region applies. The distribution of pseudo-$\ETm$ in
the control regions is used to estimate the distribution of $\ETm$ expected from the $\Zvvjets$
and $\Wlvjets$ backgrounds in the signal region.

The dimuon control region is defined using the signal region selection criteria without the muon veto. Exactly two
isolated muons with opposite charge, $\pt^{\mu_{1}},\,\pt^{{\mu}_{2}}>20,\,10$\GeV and an invariant mass in the range 60--120\GeV are required.
As the decay branching fraction of $\mathcal{B}(\Zmm)$ is approximately six times smaller than that
to neutrinos, the resulting statistical uncertainty in the \Zvvjets~background
becomes a dominant systematic uncertainty at large values of \ETm.  A
complementary approach is to use events in data that have a high-$\pt$ photon
recoiling against jets to further constrain the \Zvvjets~\cite{zjetspapers}.
This is advantageous since the production
cross section of \phojets~is roughly a factor of three times that of the \Zvvjets, yielding thereby a
smaller statistical uncertainty in the predicted background. However, the theoretical uncertainties
associated with the translation of the kinematics in $\phojets$ events to that of $\Zvvjets$ events
are significant. A combination of both photon and dimuon control regions is used
to maximally constrain the $\Zvvjets$ background.

The photon control region consists of events that are selected by a trigger requiring an
isolated photon with $\pt>150$\GeV~\cite{Khachatryan:2015iwa}.
The selected events are required to have at least one photon with $\pt > 170$\GeV
and $\abs{\eta}<2.5$, identified using a medium efficiency selection criterion~\cite{Khachatryan:2015iwa}.
Photons in the ECAL transition
region, $1.44 <\abs{\eta}< 1.56$ are excluded. All other kinematic selections are the same as those used for the signal region. The purity of the selection has been measured and is
used to estimate the contributions from other backgrounds in the photon control region~\cite{Khachatryan:2015iwa}.

To estimate the $\Wlvjets$ background, a single muon control region is defined by selecting events
with exactly one muon with $\pt>20$\GeV. Additionally, the transverse
mass, calculated as $m_{\mathrm{T}}=\sqrt{\smash[b]{2\ETm \pt^{\mu} (1-\cos \phi)}}$, where $\phi$
is the azimuthal angle between $\ptvecmiss$ and the direction of the muon momentum, is required to be in the range 50--100\GeV.

The \ETm spectra of the $\Vjets$ backgrounds are determined through the
use of a binned likelihood fit to the data in all the bins of the three control regions.
The expected number of events $N_{i}$ in a given bin $i$ of pseudo-\ETm is defined
as $N^{\Z_{\mu\mu}}_{i}=
{{\mu^{\Zvv}_{i}}}/{R^{\Z}}$ and $N^{\gamma}_{i}=
{{\mu^{\Zvv}_{i}}}/{R^{\gamma}}$ for the dimuon and photon control
regions, and $N^{\PW}_{i} = {{\mu^{\Wlv}_{i}}}/{R^{\PW}_{i}}$ for the
single muon control region. The $\mu_{i}^{\Zvv}$ and
$\mu_{i}^{\Wlv}$ terms are free parameters of the likelihood representing
the yields of $\Zvvjets$ and $\Wlvjets$ in each bin of the signal regions. The
additional terms $R^{\PW}_{i}$, $R^{\Z}_{i}$, and $R^{\gamma}_{i}$ denote factors that
account for the extrapolation of specific backgrounds from the signal region to
control regions. The likelihood function for a particular event category is
given by
\begin{equation}\begin{aligned}
\mathcal{L}(\boldsymbol{\mu}^{\Zvv},\boldsymbol{\mu}^{\Wlv},\boldsymbol{\alpha},\boldsymbol{\beta})=&
\prod_{i} \mathrm{Poisson}\left(d^{\gamma}_{i}|\left[B^{\gamma}_{i}(\boldsymbol{\alpha}) +\frac{ \mu^{\Zvv}_{i}}{R^{\gamma}_{i}(\boldsymbol{\beta})} \right] \right) \\ &\times
\prod_{i} \mathrm{Poisson}\left(d^\Z_{i} |\left[B^\Z_{i}(\boldsymbol{\alpha})  +\frac{ \mu^{\Zvv}_{i}}{R^\Z_{i} (\boldsymbol{\beta})} \right] \right) \\ &\times
\prod_{i} \mathrm{Poisson}\left(d^{\PW}_{i} |\left[B^{\PW}_{i}(\boldsymbol{\alpha})  +\frac{ \mu^{\Wlv}_{i}}{R^{\PW}_{i} (\boldsymbol{\beta})} \right] \right),
\end{aligned}\label{eqn:candclh} \end{equation}
where $d^{\gamma}_{i}$, $d^\Z_{i}$, and $d^{\PW}_{i}$ are the observed number of events in each bin, $i$, of
the photon, dimuon, and single muon control regions and $\mathrm{Poisson}(x|y)=y^{x}\re^{-y}/x!$. The terms
$\boldsymbol{\alpha,\beta}$ denote constrained nuisance parameters, which model systematic uncertainties
in the translation from the pseudo-$\ETm$ distributions in the control regions of a particular event category
to the $\ETm$ distribution in the corresponding signal region. The expected contributions
from other background processes in the photon, dimuon and single muon control regions
are denoted $B^{\gamma}_{i}$, $B^{\Z}_{i}$, and $B^{\PW}_{i}$ in Eq. (~\ref{eqn:candclh}), respectively.

The factors $R^\Z_{i}$ account for the ratio of $\mathcal{B}(\Zvv)/\mathcal{B}(\Zmm)$
and the muon efficiency times acceptance in the dimuon control region, while
$R^{\gamma}_{i}$ account for the ratio of differential cross sections between
the $\Zjets$ and $\phojets$ processes and the efficiency times acceptance of the
photon selection for the $\phojets$ control region. The differential cross sections
of photon and Z production are corrected using NLO
k-factors obtained from a comparison of their $\pt$ distributions in events generated
with {\MADGRAPH{}5\_a\MCATNLO 2.2.2}~\cite{amcatnlo}, to the
distributions produced at LO. These k-factors are propagated to the factors
$R^{\gamma}_{i}$ to account for NLO QCD effects.

\begin{figure}[hbtp]
\centering
{\includegraphics[width=0.48\textwidth]{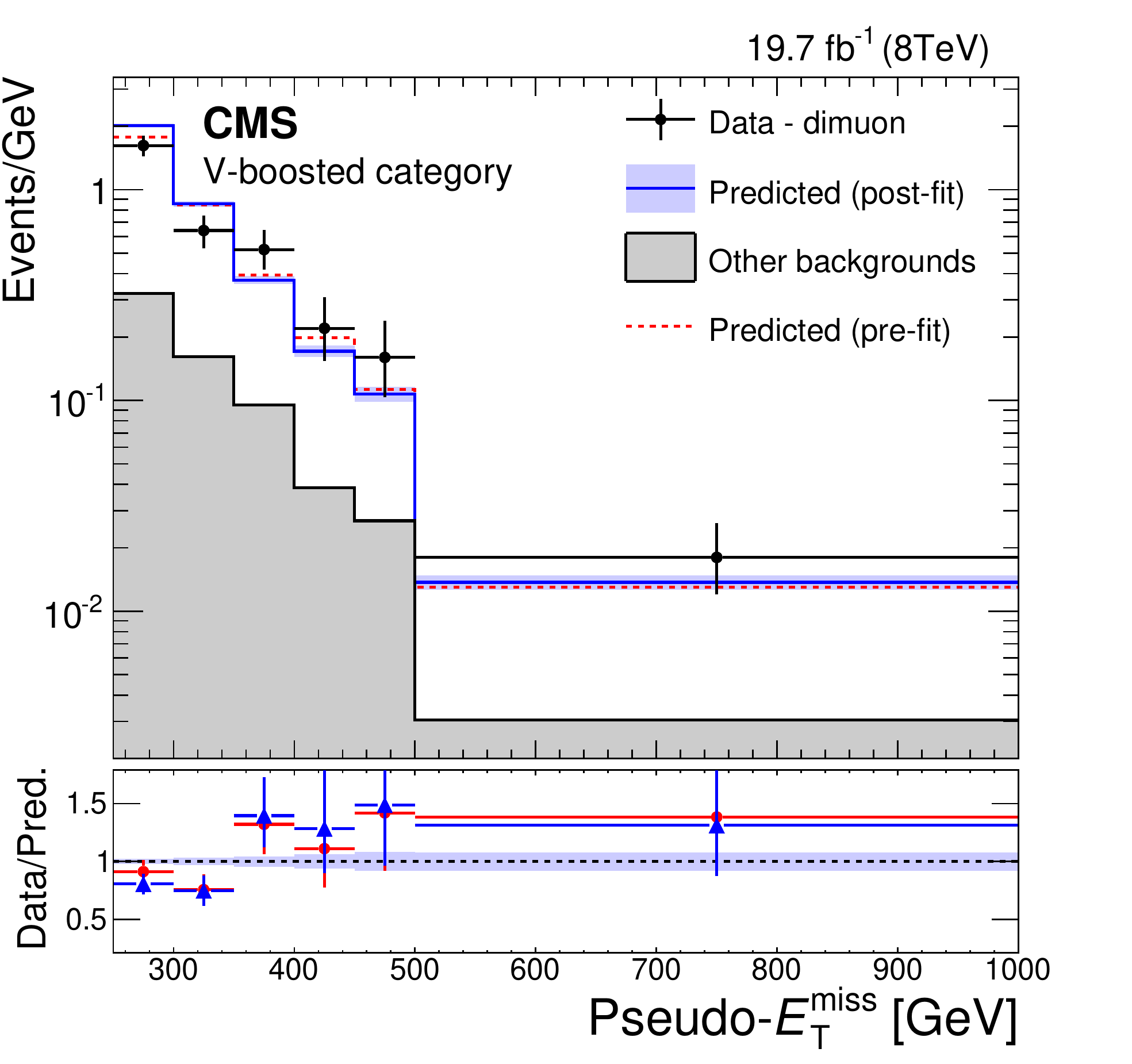}}
{\includegraphics[width=0.48\textwidth]{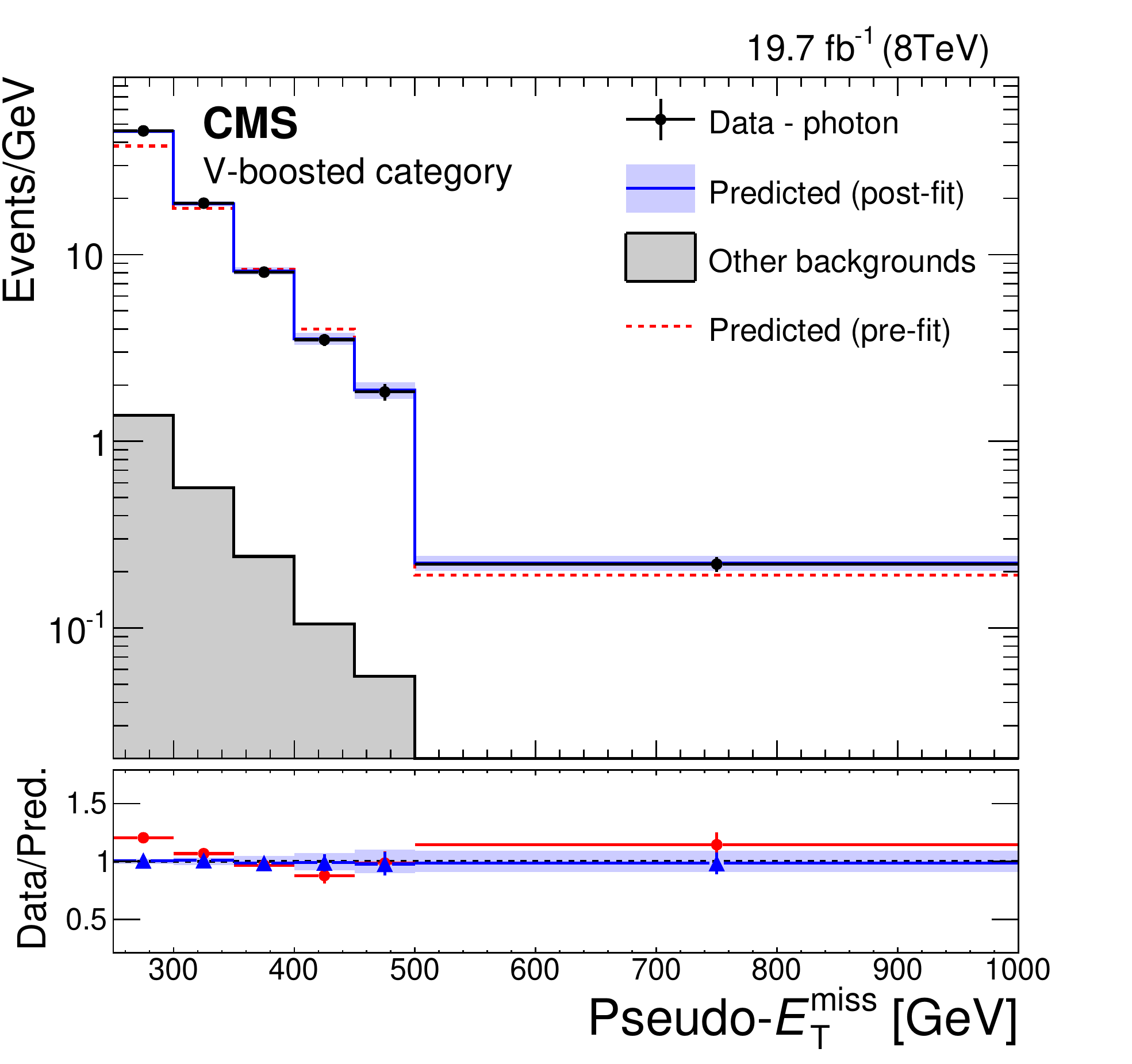}}\\
{\includegraphics[width=0.48\textwidth]{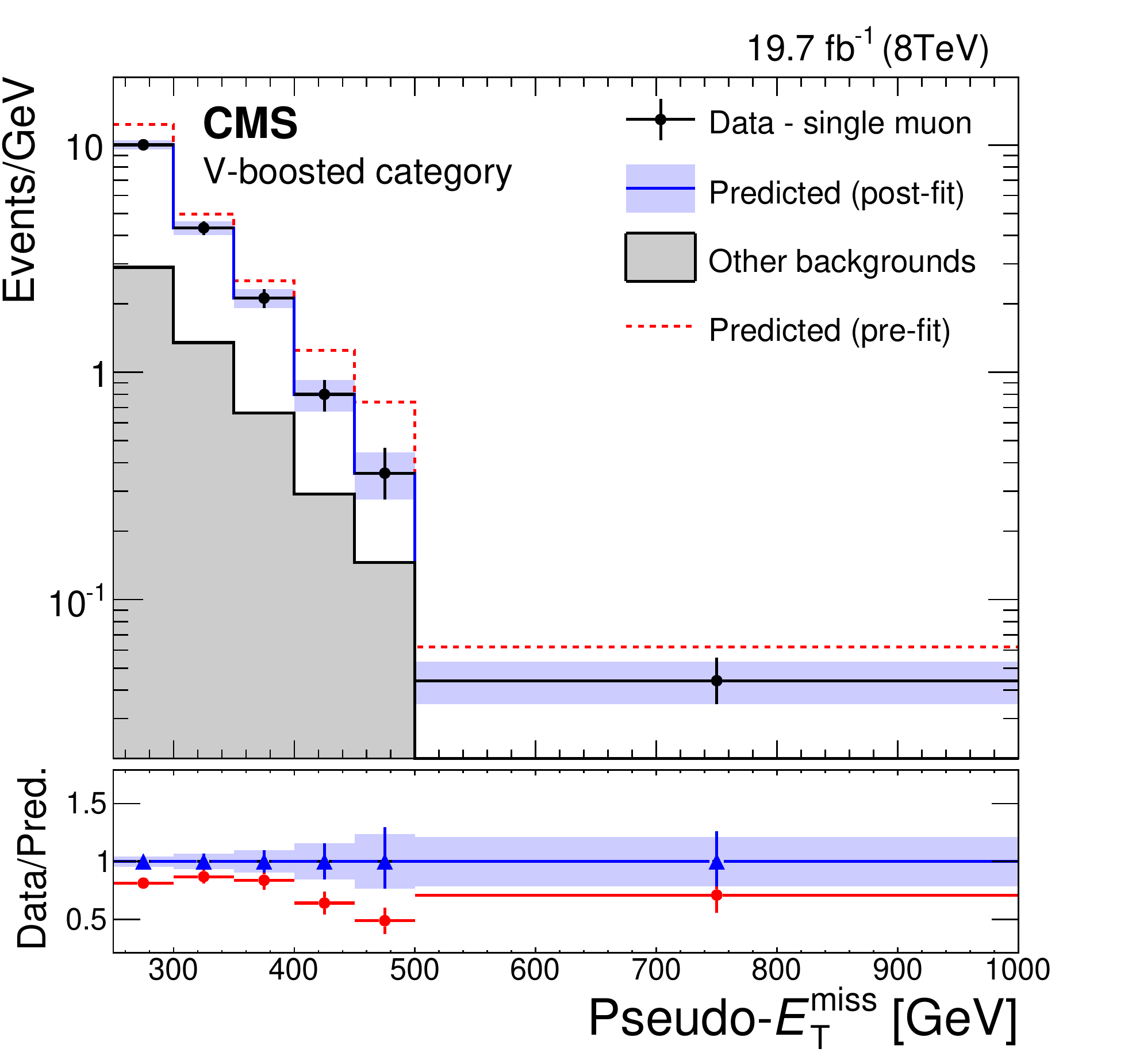}}
\caption{
Predicted and observed pseudo-\ETm distributions in the
dimuon (top--left), photon (top--right), and single muon (bottom) control regions, before and
after performing the simultaneous likelihood fit to the data in the control regions, for the V-boosted category.
The predictions for the distributions before fitting to the control region data (pre-fit), and after (post-fit)
are shown as the dashed red and solid blue lines, respectively.
The red circles in the lower panels show the ratio of the observed data to the pre-fit predictions, while the
blue triangles show the ratio to the post-fit predictions.
The horizontal bars on the data points indicate the width of the bin that is centred at that point.
The filled bands around the post-fit prediction
indicate the combined statistical and systematic uncertainties from the fit.
\label{fig:combined_fit_result_VB} }
\end{figure}

\begin{figure}[hbtp]
\centering
{\includegraphics[width=0.48\textwidth]{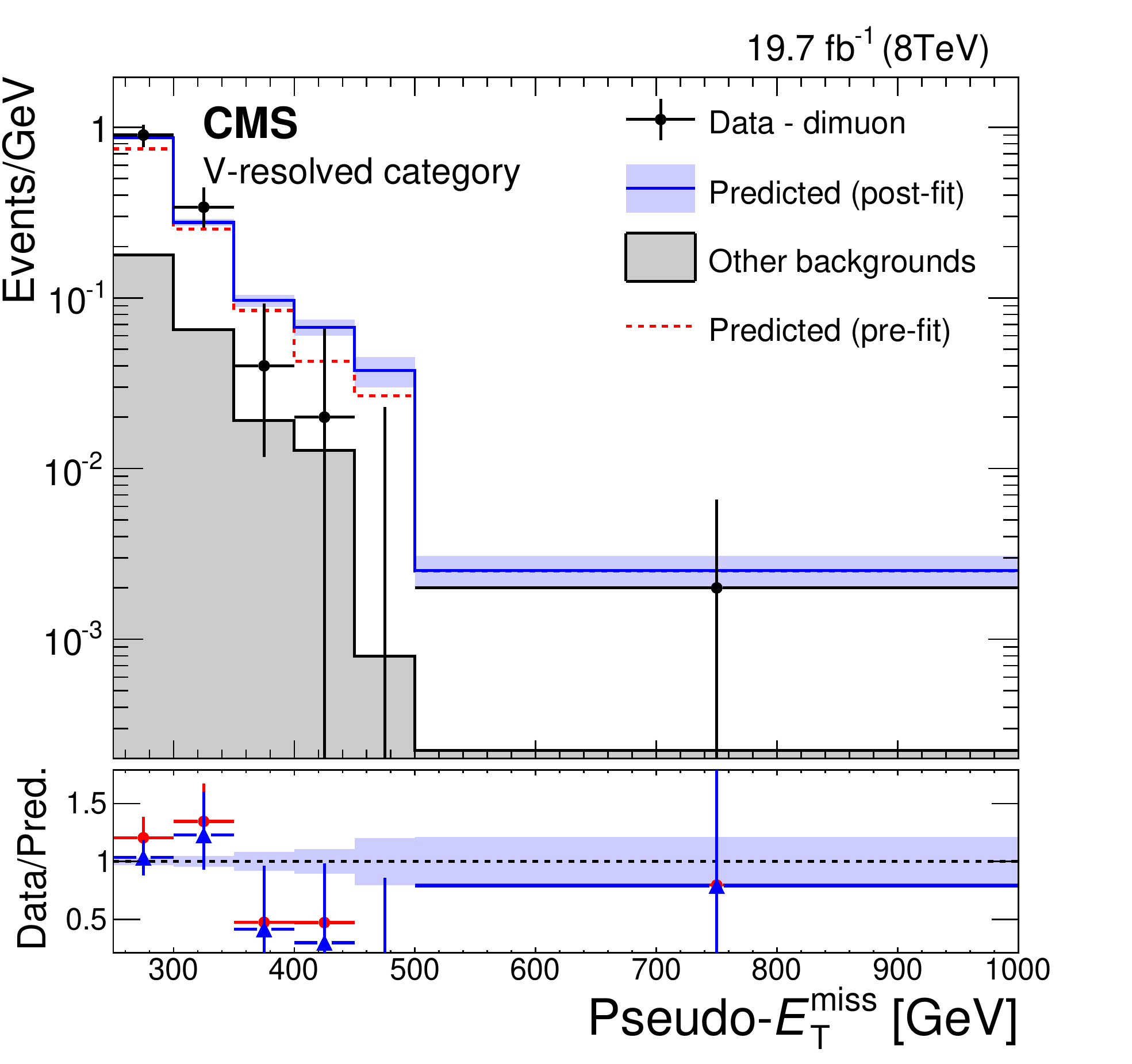}}
{\includegraphics[width=0.48\textwidth]{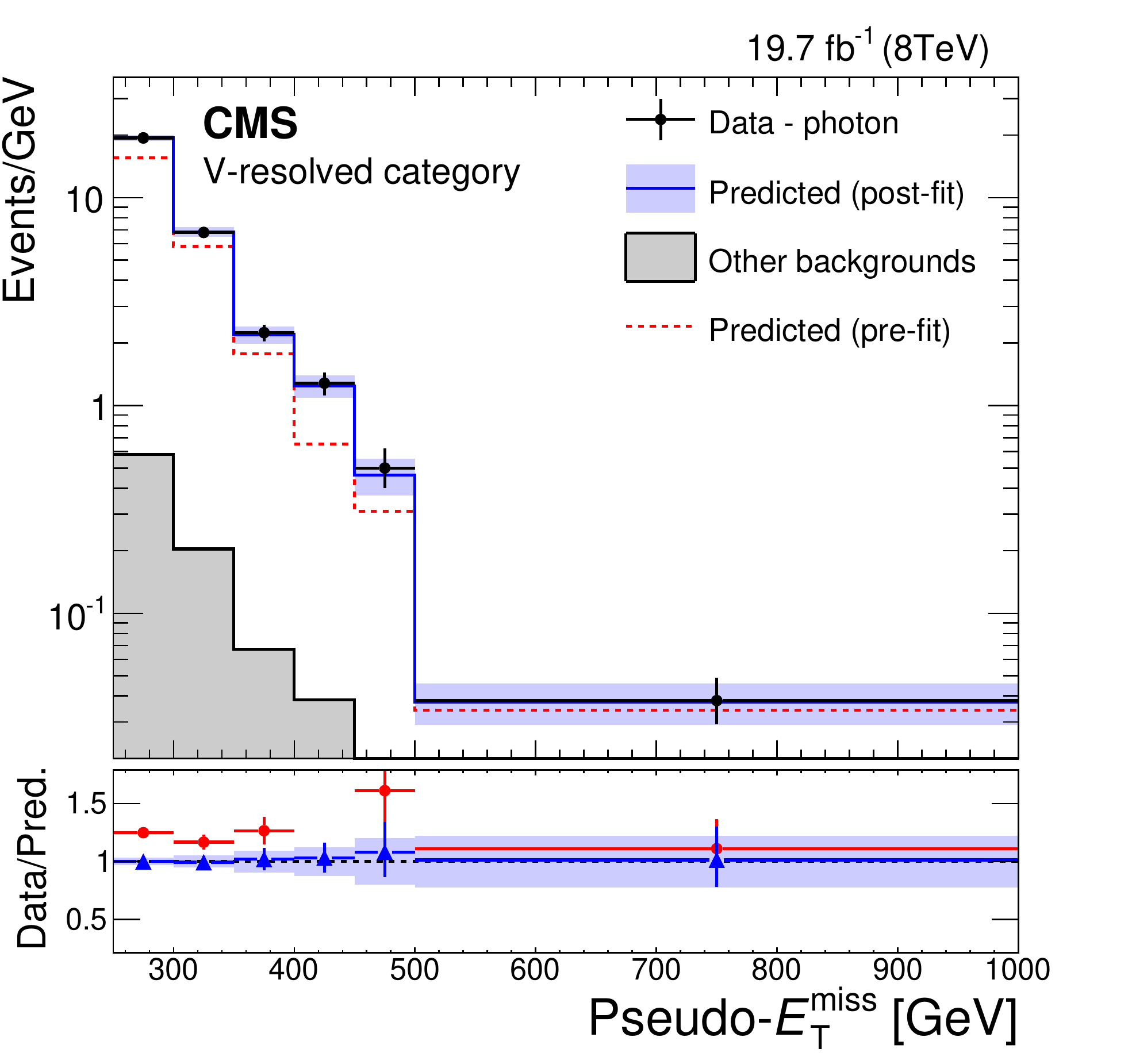}}\\
{\includegraphics[width=0.48\textwidth]{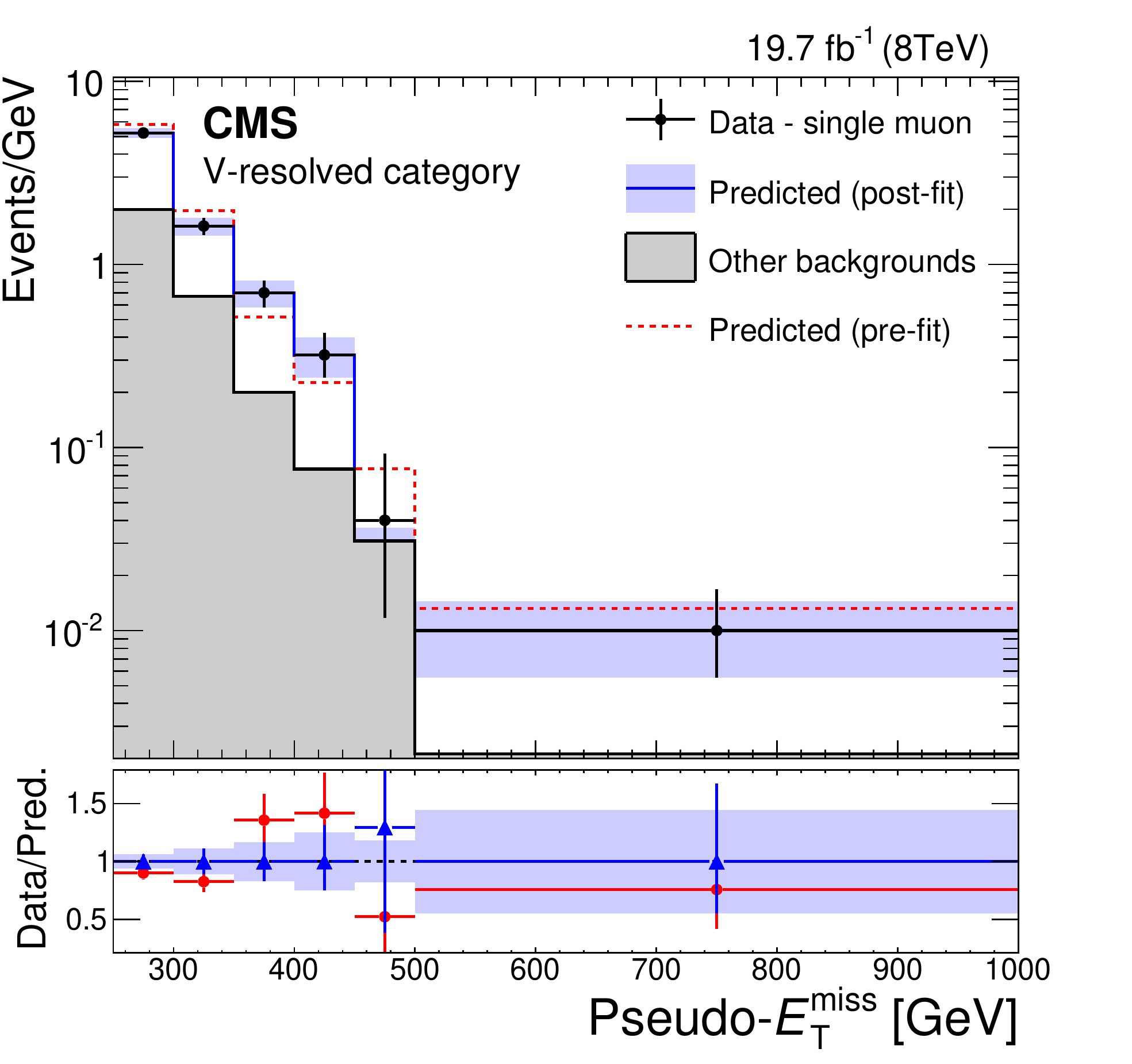}}
\caption{
Predicted and observed pseudo-\ETm distributions in the
dimuon (top--left), photon (top--right), and single muon (bottom) control regions, before and
after performing the simultaneous likelihood fit to the data in the control regions, for the V-resolved category.
The predictions for the distributions before fitting to the control region data (pre-fit), and after (post-fit)
are shown as the dashed red and solid blue lines, respectively.
The red circles in the lower panels show the ratio of the observed data to the pre-fit predictions, while the
blue triangles show the ratio to the post-fit predictions.
The horizontal bars on the data points indicate the width of the bin that is centred at that point.
The filled bands around the post-fit prediction
indicate the combined statistical and systematic uncertainties from the fit.
\label{fig:combined_fit_result_VR} }
\end{figure}

\begin{figure}[hbtp]
\centering
{\includegraphics[width=0.48\textwidth]{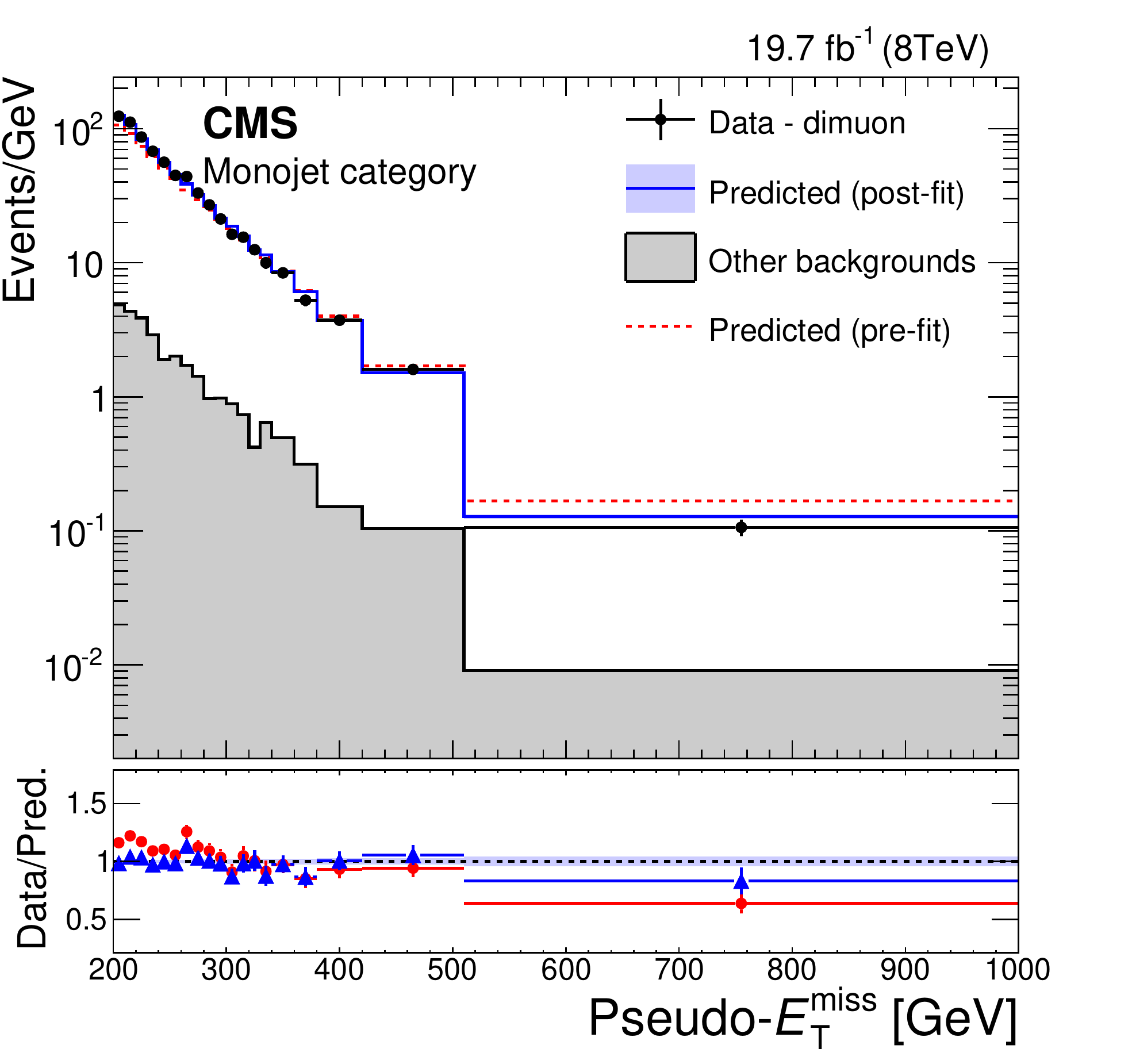}}
{\includegraphics[width=0.48\textwidth]{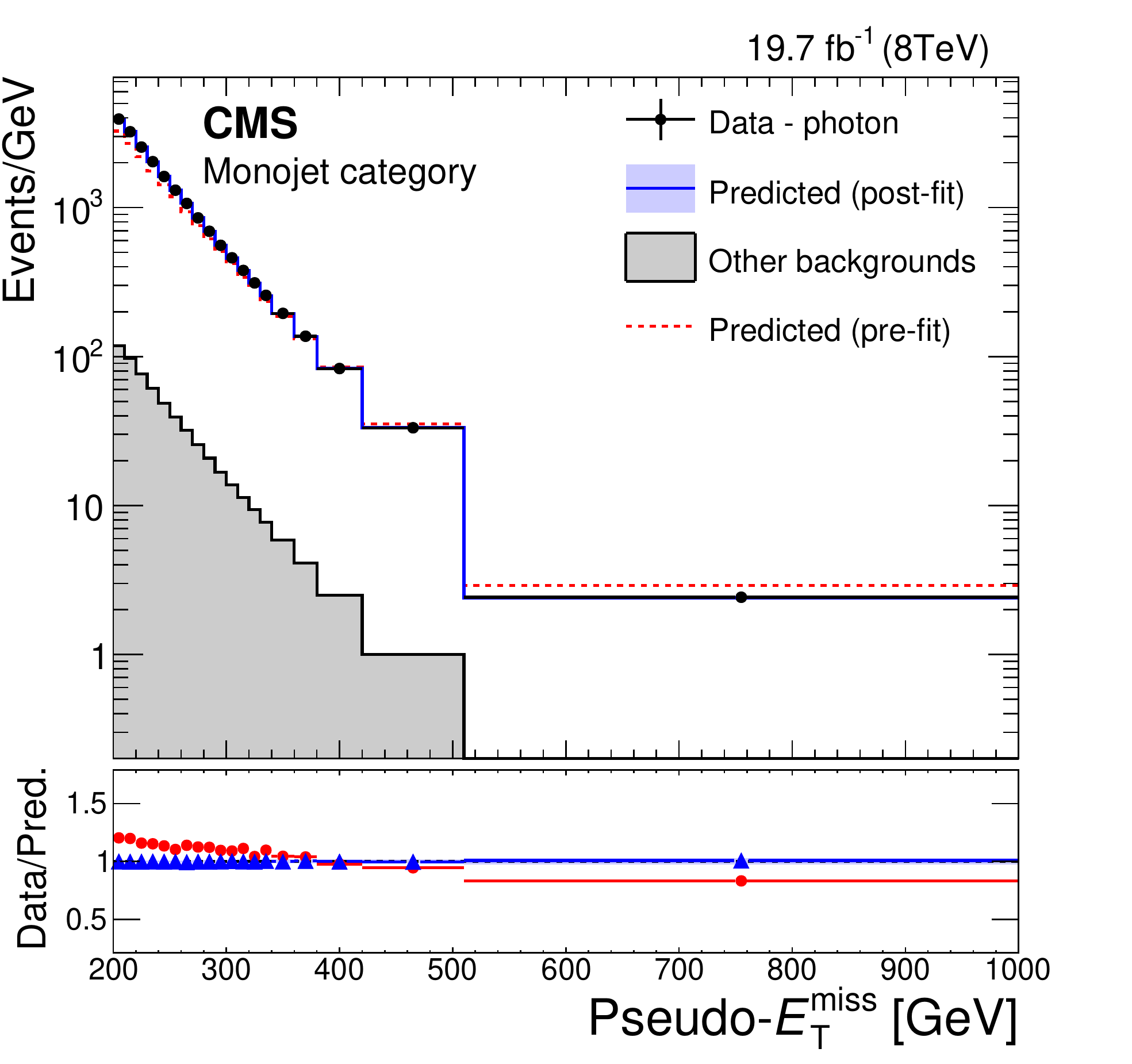}}\\
{\includegraphics[width=0.48\textwidth]{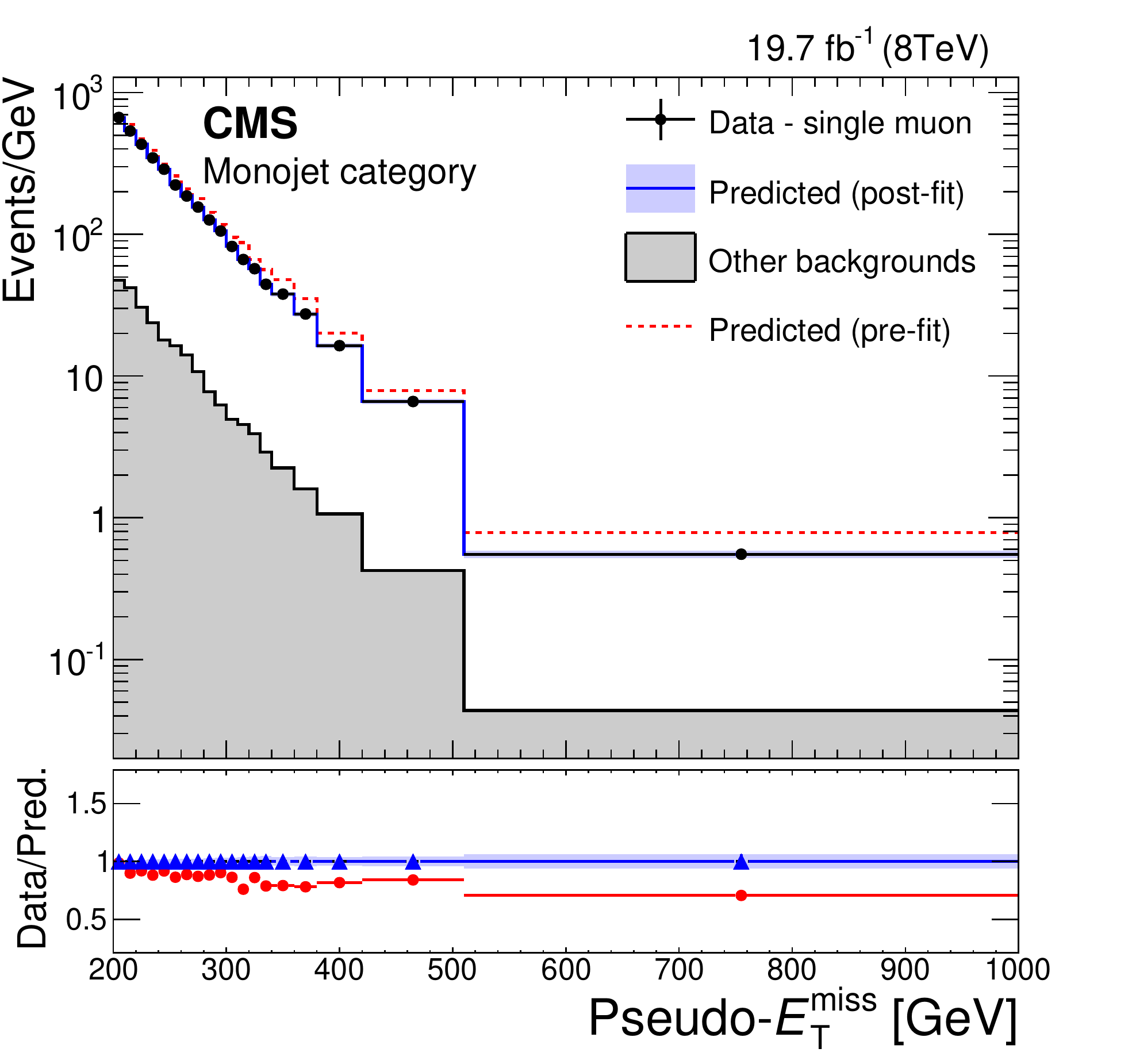}}
\caption{
Predicted and observed pseudo-\ETm distributions in the
dimuon (top--left), photon (top--right), and single muon (bottom) control regions, before and
after performing the simultaneous likelihood fit to the data in the control regions, for the monojet category.
The predictions for the distributions before fitting to the control region data (pre-fit), and after (post-fit)
are shown as the dashed red and solid blue lines, respectively.
The red circles in the lower panels show the ratio of the observed data to the pre-fit predictions, while the
blue triangles show the ratio to the post-fit predictions.
The horizontal bars on the data points indicate the width of the bin that is centred at that point.
The filled bands around the post-fit prediction
indicate the combined statistical and systematic uncertainties from the fit.
\label{fig:combined_fit_result_MJ} }
\end{figure}

Systematic uncertainties are modelled as constrained nuisance parameters that allow variation of the factors
$R^{\gamma}_{i}$, $R^{\Z}_{i}$ and $R^{\PW}_{i}$ in the fit. These include theoretical uncertainties in the photon to Z
differential cross section ratio from renormalization and factorization scale
uncertainties, which amount to 8\% each across the relevant boson \pt range.
These uncertainties are conservative in that they are estimated by taking the maximum difference in the ratio derived from varying
each scale by a factor of two, independently for the two processes, thereby ignoring any cancellation of the scale uncertainties.
Electroweak corrections are not accounted for in
the simulation. Additional k-factors are applied as a function
of the boson (Z or $\gamma$) \pt, to account for higher order electroweak effects, which are around 15\%
for a boson \pt around 1\TeV~\cite{Kuhn:2005gv}. The full correction is taken as an uncertainty in the
ratio. A conservative choice is made in assuming this
uncertainty to be uncorrelated across bins of \ETm. The
uncertainties in the muon selection efficiency, photon selection efficiency, and
photon purity are included and fully correlated across the control regions for the three event categories. The results
of the fit to the data in the control regions for the V-boosted, V-resolved, and monojet categories are shown in
Figs.~\ref{fig:combined_fit_result_VB},~\ref{fig:combined_fit_result_VR}, and~\ref{fig:combined_fit_result_MJ}, respectively.

The remaining backgrounds are expected to be much smaller than those from $\Vjets$
and are estimated directly from simulation. Shape and normalization
systematic uncertainties from the hadronic recoil corrections applied to these
backgrounds are included and account for uncertainties in the jet energy scale and
resolution.
Systematic uncertainties related to the V-tagging efficiency of both of the V-tagged categories
are included for the top and diboson backgrounds, which allow for migration of events between the
three categories. The systematic uncertainty is roughly 10\% in the V-resolved category, which
allows for the disagreement between data and MC observed in the MVA distribution (Fig.~\ref{fig:vtagger}) and
10\% in the V-boosted category, which allows for the uncertainty in the measurement of the selection
efficiency using ttbar events in data~\cite{Khachatryan:2014vla}.
A systematic uncertainty of 4\% is included
for the top quark backgrounds normalization because of the uncertainty in the b tagging efficiency for
the b jet veto in the V-resolved category~\cite{CMS-PAS-BTV-13-001}. Systematic uncertainties of 7\% and 10\%
are included in the normalizations of the top quark~\cite{Khachatryan:2015oqa} and
diboson~\cite{tagkey2015250,Khachatryan:2015sga} backgrounds, respectively, to account for the uncertainty in their
cross sections in the relevant kinematic phase-space. The top quark and diboson backgrounds have been studied
separately using dedicated control regions in data to validate these
systematic uncertainties. A systematic uncertainty of 50\% is included in the expected
contribution from QCD multijet events. This uncertainty was obtained by taking the largest differences
observed between data and simulation in events selected by inverting the requirement on $\Delta\phi(\ptvecmiss,\mathrm{j})$.
Finally, a systematic uncertainty of 2.6\% in the integrated luminosity measurement~\cite{lumi} is included in the normalization all of the backgrounds obtained from simulation.

The expected yields in each bin of \ETm from all SM backgrounds,
after the fit to the data in the control regions, are given in
Tables~\ref{tab:bkgboosted}, \ref{tab:bkgresolved}, and~\ref{tab:bkgmonojet} for the V-boosted,
V-resolved, and monojet signal region, respectively.
The uncertainties represent the sum in quadrature of the effects of all the relevant
sources of systematic uncertainty in each bin of \ETm.
The correlations between the \ETm bins, resulting from the fit to the control regions, for each of the three event categories are
shown in Figs.~\ref{fig:correlation_VB},~\ref{fig:correlation_VR}, and~\ref{fig:correlation_MJ} of the supplementary material in Appendix~\ref{app:suppMat}.

\begin{table}[htbp]
  \centering
    \topcaption{Expected yields of the SM processes and their uncertainties per bin for the V-boosted category after the fit to the control regions.
      \label{tab:bkgboosted}}
\cmsTable{\begin{tabular}{l|c|c|c|c|c|c|c}
	\hline
	$\ETm$\,(\GeVns{}) & Obs.  & Z($\to \nu\nu$)+jets &  W($\to \ell\nu$)+jets & Top quark & Dibosons &  Other  & Total Bkg. \\
\hline

      250--300 & 1073  &    683$\pm$40  &    279$\pm$33  &     35.4$\pm$3.7  &    103$\pm$15  &       2.5$\pm$0.1  &   1103$\pm$63  \\

      300--350 & 453  &    271$\pm$23  &    114$\pm$20  &     12.7$\pm$1.3  &     46.5$\pm$6.9  &     0.7$\pm$0.1  &    446$\pm$34  \\

      350--400 & 160  &    118$\pm$13  &     38.3$\pm$8.7  &      5.6$\pm$1.0  &     22.2$\pm$3.3  &  0.2$\pm$0.1  &    184$\pm$18  \\

      400--450 & 81  &     49.7$\pm$7.3  &      9.8$\pm$3.4  &      1.5$\pm$0.8  &     11.0$\pm$1.8   & $<$0.1     &     72$\pm$29  \\

      450--500 & 30  &     31.2$\pm$6.1  &      5.0$\pm$2.6  &      0.5$\pm$0.1  &      7.4$\pm$1.1   & $<$0.1     &     44.3$\pm$6.6  \\

      500--1000 & 39  &     39.8$\pm$7.8  &      6.4$\pm$3.4  &      0.2$\pm$0.0  &      7.8$\pm$1.1  & $<$0.1     &     54.3$\pm$8.5  \\

\hline
    \end{tabular}}
\end{table}
\begin{table}
    \topcaption{Expected yields of the SM processes and their uncertainties per bin for the V-resolved category after the fit to the control regions.
      \label{tab:bkgresolved}}
    \cmsTable{\begin{tabular}{l|c|c|c|c|c|c|c}
	\hline
	$\ETm$ (\GeVns{}) & Obs.  & Z($\to \nu\nu$)+jets &  W($\to \ell\nu$)+jets & Top quark & Dibosons &  Other  & Total Bkg. \\
\hline

      250--300 & 617  &    298$\pm$36  &    166$\pm$26  &     55.4$\pm$4.7  &     27.9$\pm$1.6       &     36$\pm$17  &    587$\pm$48  \\

      300--350 & 211  &     98$\pm$14  &     41$\pm$10  &     15.2$\pm$1.5  &      9.6$\pm$0.3       &     19.2$\pm$6.6  &    170$\pm$18  \\

      350--400 & 79  &     31.1$\pm$7.0  &     21.5$\pm$8.9  &      5.5$\pm$0.7  &      3.2$\pm$0.3  &      8.2$\pm$2.3  &     62$\pm$12  \\

      400--450 & 20  &     20.1$\pm$6.4  &     14.5$\pm$8.5  &      1.5$\pm$0.2  &      0.6$\pm$0.3  &      3.0$\pm$0.7  &     38$\pm$11  \\

      450--500 & 16  &      6.1$\pm$2.7  &      1.0$\pm$2.6  &      1.0$\pm$0.4  &      0.4$\pm$0.1   &     1.0$\pm$0.2  &     8.5$\pm$3.6  \\

      500--1000 & 17  &      6.9$\pm$3.0  &      2.6$\pm$1.7  &      0.3$\pm$0.2  &      0.5$\pm$0.0  &     0.3$\pm$0.1  &     11.6$\pm$3.5  \\

\hline
\end{tabular}}
\end{table}

\begin{table}[htbp]
\centering
    \topcaption{Expected yields of the SM processes and their uncertainties per bin for the monojet category  after the fit to the control regions.
      \label{tab:bkgmonojet}}
    \cmsTable{\begin{tabular}{c|c|c|c|c|c|c|c}
	\hline
	$\ETm$ (\GeVns{}) & Obs.  & Z($\to \nu\nu$)+jets &  W($\to \ell\nu$)+jets & Top quark & Dibosons &  Other  & Total Bkg. \\

\hline

      200--210 & 17547  &  10740$\pm$270  &   6770$\pm$320  &    132$\pm$11  &    135$\pm$14        &    93.4$\pm$16.9  & 17870$\pm$600  \\

      210--220 & 14303  &   9230$\pm$230  &   4990$\pm$240  &    104$\pm$13  &    112$\pm$11  	    &    63.7$\pm$6.7  &  14500$\pm$610  \\

      220--230 & 11343  &   7320$\pm$190  &   3830$\pm$170  &     82.1$\pm$7.3  &     95.1$\pm$9.6  &    39.4$\pm$2.4  &  11370$\pm$400  \\

      230--240 & 8961  &   5730$\pm$170  &   3020$\pm$160  &     62.0$\pm$5.8  &     77.9$\pm$8.6  &     29.0$\pm$1.0  &   8920$\pm$400  \\

      240--250 & 6920  &   4680$\pm$150  &   2470$\pm$140  &     46.6$\pm$4.4  &     61.0$\pm$6.1  &     19.6$\pm$0.5 &    7280$\pm$330  \\

      250--260 & 5582  &   3700$\pm$140  &   1860$\pm$120  &     34.2$\pm$3.7  &     50.1$\pm$4.9  &     14.6$\pm$0.4  &   5660$\pm$370  \\

      260--270 & 4517  &   3290$\pm$130  &   1580$\pm$110  &     27.7$\pm$2.3  &     39.7$\pm$4.2  &     10.3$\pm$0.2  &   4940$\pm$320  \\

      270--280 & 3693  &   2570$\pm$110  &   1101$\pm$71  &     25.0$\pm$3.1  &     33.5$\pm$3.4  &      6.3$\pm$0.2  &   3730$\pm$160  \\

      280--290 & 2907  &   2085$\pm$89  &    934$\pm$71  &     17.8$\pm$1.9  &     28.1$\pm$3.0  &       5.5$\pm$0.1  &   3070$\pm$180 \\

      290--300 & 2406  &   1721$\pm$85  &    754$\pm$58  &     15.0$\pm$3.6  &     21.9$\pm$2.7  &       4.2$\pm$0.1  &   2520$\pm$170  \\

      300--310 & 1902  &   1337$\pm$79  &    577$\pm$51  &      8.9$\pm$1.6  &     17.7$\pm$2.1  &       3.1$\pm$0.1  &   1940$\pm$160  \\

      310--320 & 1523  &   1182$\pm$58  &    435$\pm$43  &      5.9$\pm$2.2  &     15.5$\pm$1.8  &       2.3$\pm$0.1  &   1640$\pm$110  \\

      320--330 & 1316  &    931$\pm$53  &    371$\pm$44  &      5.2$\pm$1.3  &     11.0$\pm$1.8  &       2.1$\pm$0.1  &   1320$\pm$92  \\

      330--340 & 1065  &    804$\pm$51  &    246$\pm$29  &      4.9$\pm$1.1  &     11.9$\pm$1.8  &       1.8$\pm$0.1  &   1070$\pm$120  \\

      340--360 & 1571  &   1225$\pm$61  &    399$\pm$39  &      6.8$\pm$1.2  &     16.4$\pm$1.6  &       2.0$\pm$0.1  &   1650$\pm$110  \\

      360--380 & 1091  &    822$\pm$53  &    269$\pm$30  &      3.4$\pm$0.4  &     13.3$\pm$1.4  &      1.3$\pm$0.1  &   1110$\pm$150  \\

      380--420 & 1404  &   1036$\pm$66  &    324$\pm$30  &      5.5$\pm$0.6  &     17.1$\pm$1.7  &      1.4$\pm$0.1  &   1390$\pm$110  \\

      420--510 & 1126  &    943$\pm$70  &    267$\pm$27  &      3.9$\pm$0.8  &     15.7$\pm$1.6  &      1.3$\pm$0.1  &   1240$\pm$140  \\

      510--1000 & 476  &    330$\pm$32  &     72$\pm$12  &      0.6$\pm$0.2  &      8.2$\pm$0.8  &      0.3$\pm$0.1  &    412$\pm$71  \\

\hline
    \end{tabular}}
\end{table}

\section{Results}
\begin{figure}[hbtp]
\centering
\includegraphics[width=0.49\textwidth]{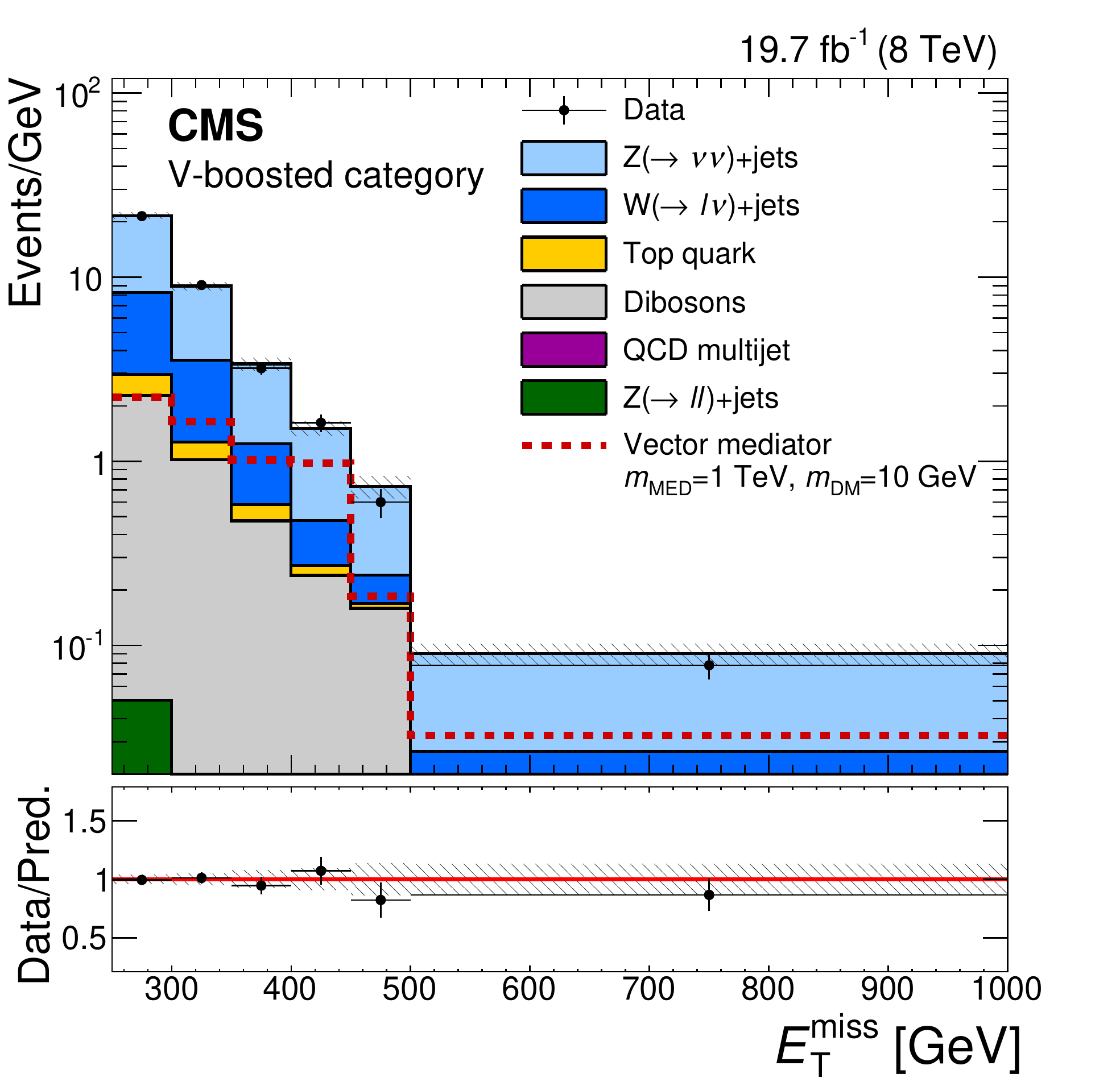}
\includegraphics[width=0.49\textwidth]{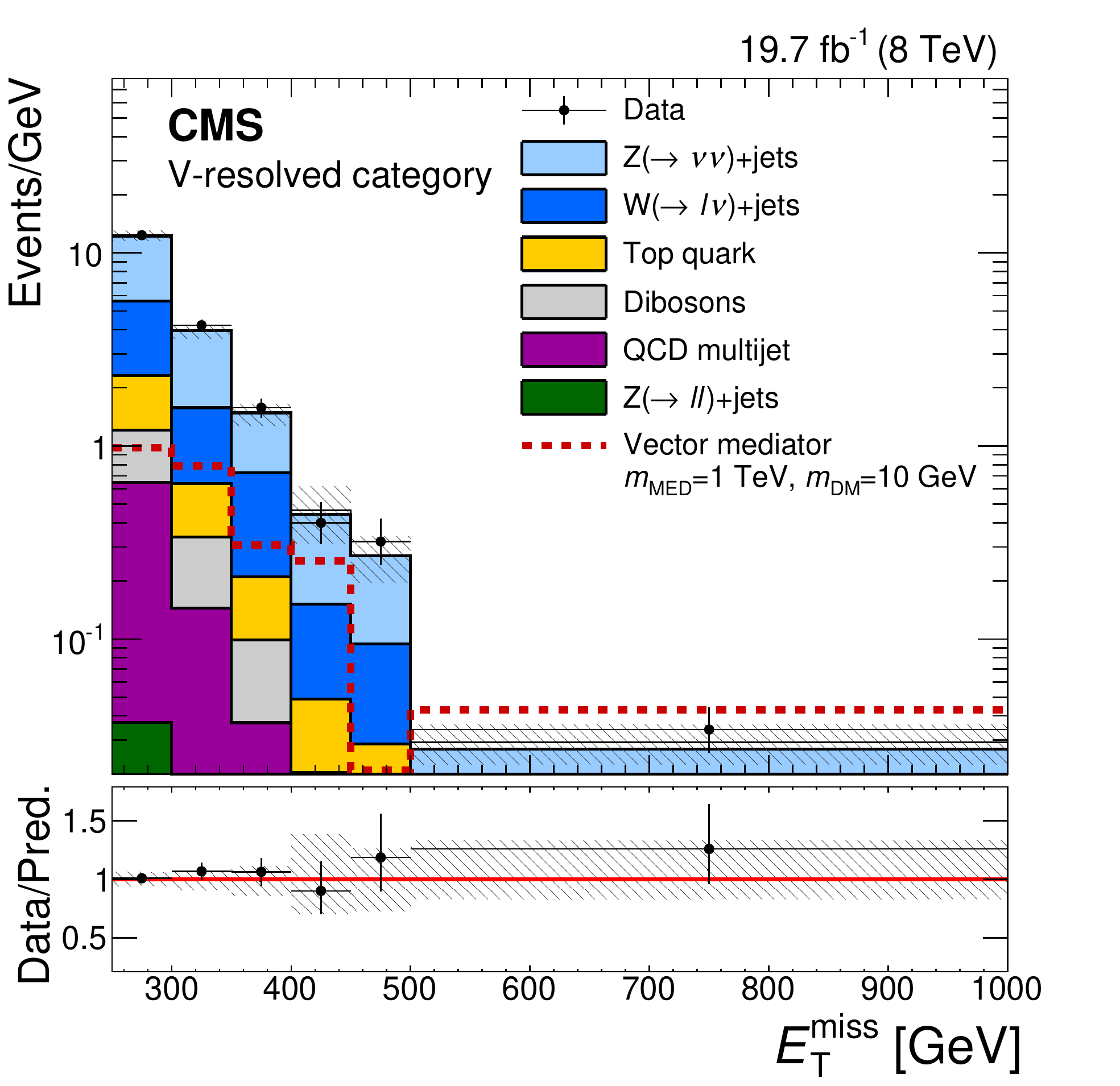}
\includegraphics[width=0.49\textwidth]{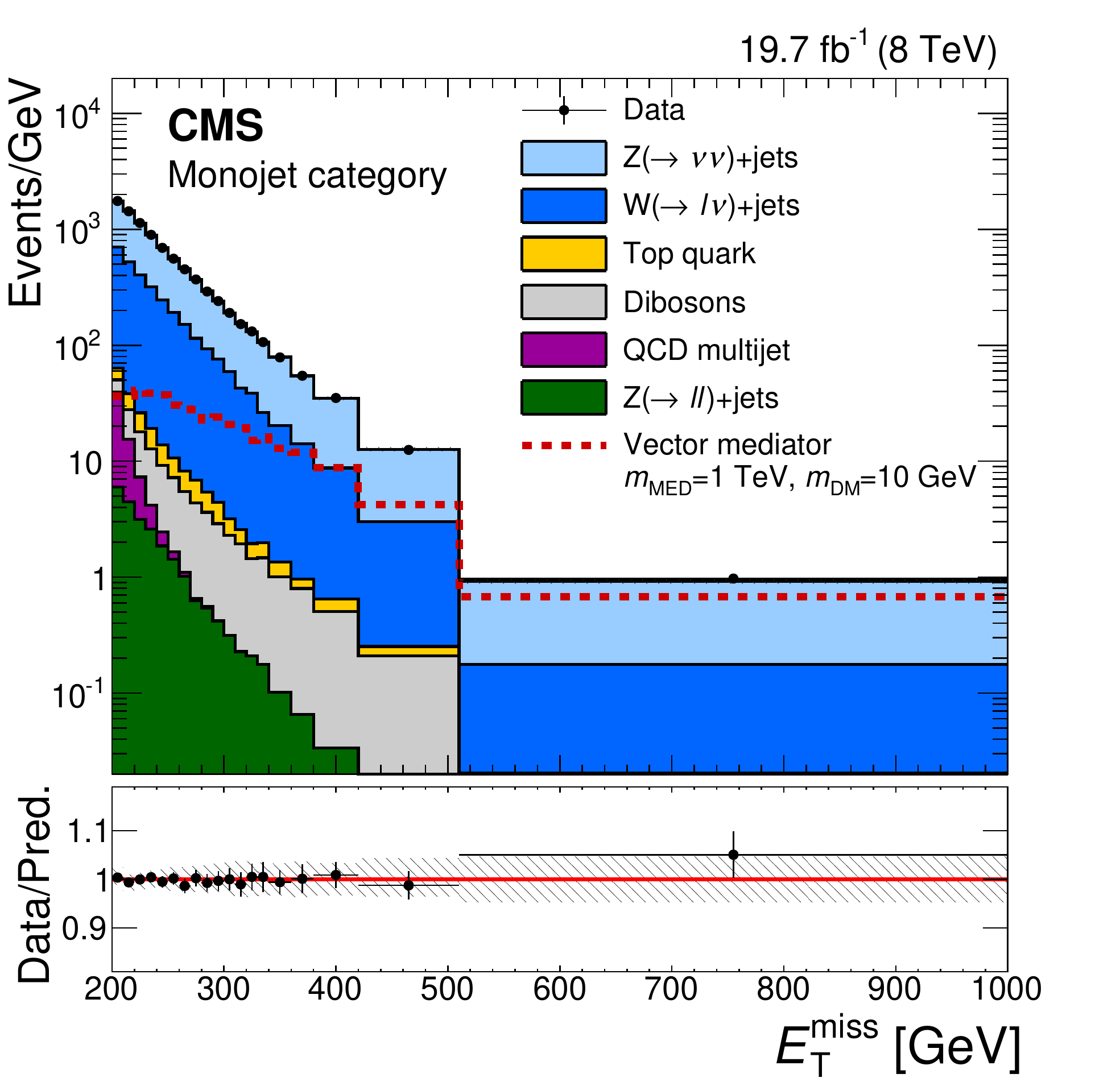}
\caption{Post-fit distributions in \ETm expected from SM backgrounds and
observed in the signal region. The expected distributions are evaluated
after fitting to the observed data simultaneously across the V-boosted (top--left),
V-resolved (top--right), and monojet (bottom) categories. The ratio of the data to the post-fit background prediction is shown in the lower panels.
The shaded bands indicate the post-fit
uncertainty in the background, assuming no signal. The horizontal bars on the data points indicate the width of the bin that is centred at that point.
The expected distribution for a signal
assuming vector mediated DM production is shown for
$m_{\mathrm{MED}}=1$\TeV and $m_{\mathrm{DM}}=10$\GeV.} \label{fig:post_fit_plots}\end{figure}

A simultaneous fit to the data in the three event category signal regions is performed. The background shapes in this second fit are
allowed to vary within their uncertainties, which are propagated from the fit to the control region data, described in the previous section,
accounting for correlations between the control region fit parameters.
The corresponding comparisons between the data and the expected backgrounds in the
\ETm distributions after this fit are shown
in Fig.~\ref{fig:post_fit_plots} for each of the three event categories.
Agreement between the data and the expected backgrounds is observed at the percent level across the three
categories. A local significance of the data in each bin is calculated by
comparing the likelihood between the background-only fit
(Fig.~\ref{fig:post_fit_plots}) and a fit in which the total expected yield of events in that bin is fixed
to the observation in data. The largest local significance
observed using this procedure is 1.9 standard deviations and corresponds to the largest \ETm
bin of the monojet category.

The results are interpreted using the set of simplified models for DM production
described in Section~\ref{sec:signals}.
Exclusion limits are set for these models using the asymptotic
CL$_\mathrm{s}$ method~\cite{Cowan:2010js,CMS-NOTE-2011-005,cls} with a
profile likelihood ratio as the test statistic, in which systematic uncertainties in the signal
and background models are modelled as constrained nuisance parameters. For each signal hypothesis
tested, upper limits are placed on the ratio of the signal yield to that predicted by the simplified model,
denoted as $\mu$. Limits are presented in
terms of excluded regions in the $m_{\mathrm{MED}}-m_{\mathrm{DM}}$ plane,
assuming scalar, pseudoscalar, vector, and axial-vector mediators,
determined as the points for which $\mu>1$ is excluded at the 90\% confidence level
(CL) or above. The choice of 90\% CL exclusions is made to allow comparison with other experiments.
Limits are calculated for a set of points in the plane and then interpolated to derive exclusion contours.
In the region $m_{\mathrm{MED}}<200$\GeV, $m_\mathrm{DM} < 200$\GeV,
the limit is calculated in 10\GeV steps
in both DM and mediator masses. For the region $200 < m_{\mathrm{MED}} < 500$\GeV, $m_{\mathrm{DM}} < 500$\GeV, a spacing
of 25\GeV is used. For mediator masses larger than 500\GeV the generated signal points are separated by 100\GeV.
The expected number of signal events in each of the three event categories arising from monojet and mono-V
production for a vector and axial vector mediator with a mass of 1\TeV, a scalar mediator with a mass of 125\GeV,
and a pseudoscalar mediator with a mass of 400\GeV is shown in Table~\ref{tab:sigyields}. The yields are derived assuming
a DM mass of 1\GeV and coupling values $g_{\mathrm{DM}}=g_{\mathrm{SM}}=g_{\PQq}=1$. The sum of these
contributions in each category is used when setting limits, except in the fermionic case, for which the contribution from the
mono-V signal is ignored.

\begin{table}[htbp]
\centering
    \topcaption{Expected signal event yields in each of the three event categories for monojet and mono-V production assuming a vector, axial vector,
		pseudoscalar, or scalar mediator. The yields are determined assuming $m_{\mathrm{DM}}=1$\GeV and $g_{\mathrm{DM}}=g_{\mathrm{SM}}=g_{\PQq}=1$.\label{tab:sigyields}}
   \cmsTable{\begin{tabular}{l | c c | c c | c c}
   \hline
     						     & \multicolumn{2}{c|}{V-boosted} &  \multicolumn{2}{c|}{V-resolved} & \multicolumn{2}{c}{Monojet}  \\
     Mediator type		                    & monojet & mono-V  & monojet & mono-V & monojet  & mono-V  \\   \hline
     Vector, $m_{\mathrm{MED}}=1$\TeV               & 217     & 84.0	 & 82.0    & 26.1   & 5250     & 94.8    \\
     Axial vector, $m_{\mathrm{MED}}=1$\TeV         & 268     & 85.7    & 85.5    & 24.5   & 6030     & 93.7    \\
     Pseudoscalar, $m_{\mathrm{MED}}=400$\GeV 	    & 56.8    & 1300    & 20.8    & 100    & 2420     & 1500    \\
     Scalar, $m_{\mathrm{MED}}=125$\GeV             & 20.6    & 126     & 8.44    & 13.3   & 1060     & 196     \\
   \hline
   \end{tabular}}
\end{table}

Experimental systematic uncertainties, including jet and \ETm response and
resolution uncertainties, are included in the signal model as nuisance parameters, while the
theoretical systematic uncertainties in the inclusive cross section are instead added as
additional contours on the exclusion limits. These include the
effect of varying the renormalization and factorization scales by a factor of two, and the PDF uncertainties, which
result in 20\% and 30\% variations in the signal yield, when summed in quadrature, for the vector and axial vector,
and scalar and pseudoscalar models, respectively. These are the largest values found across the full range of the
mediator mass from 10\GeV to 3\TeV, although the variation of these uncertainties in this range is found to be small.
The same values are assumed for every signal point, thus giving a conservative estimate of the uncertainty.

\begin{figure}[htb] \centering
\includegraphics[width=0.49\textwidth]{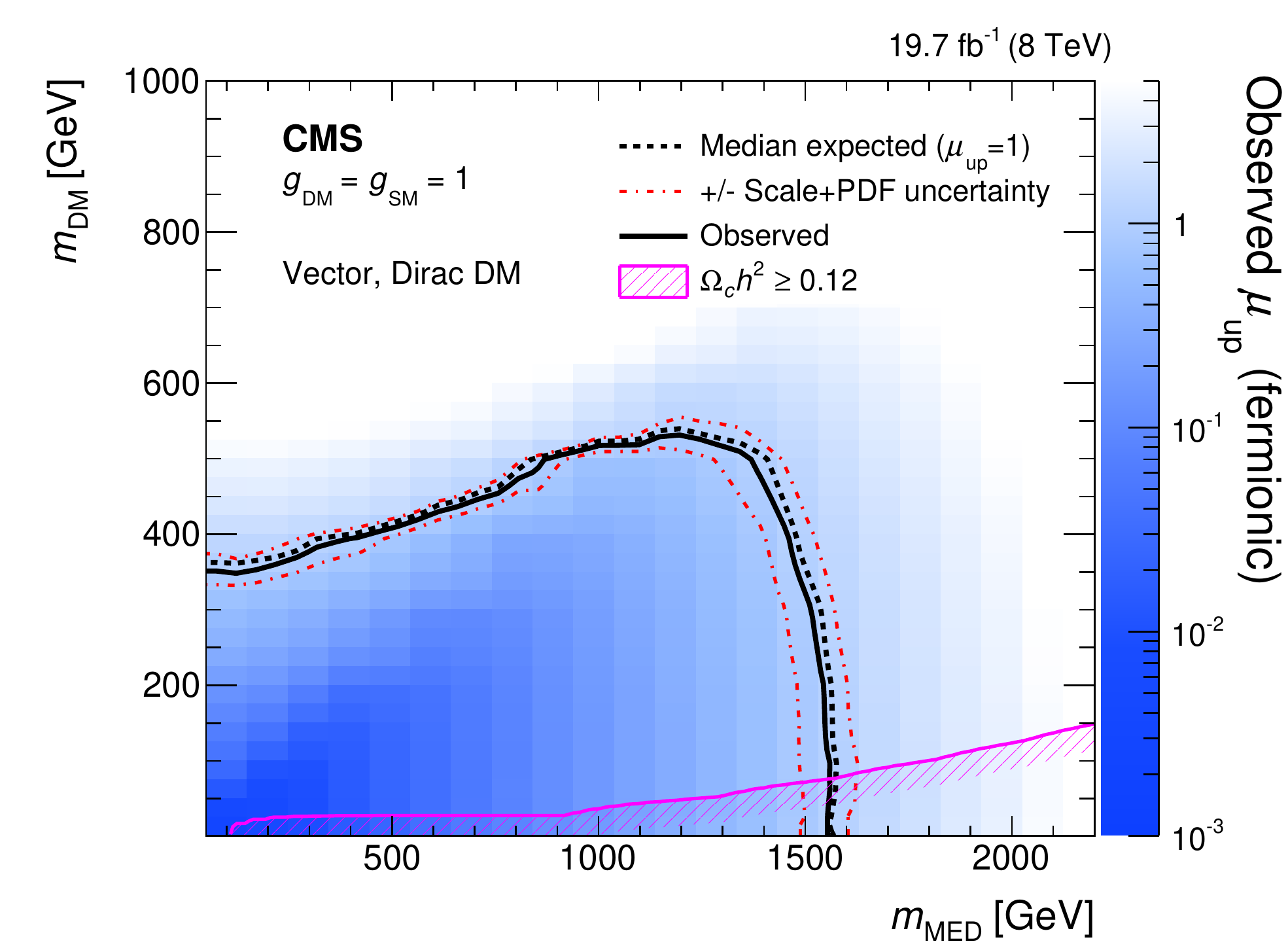}
\includegraphics[width=0.49\textwidth]{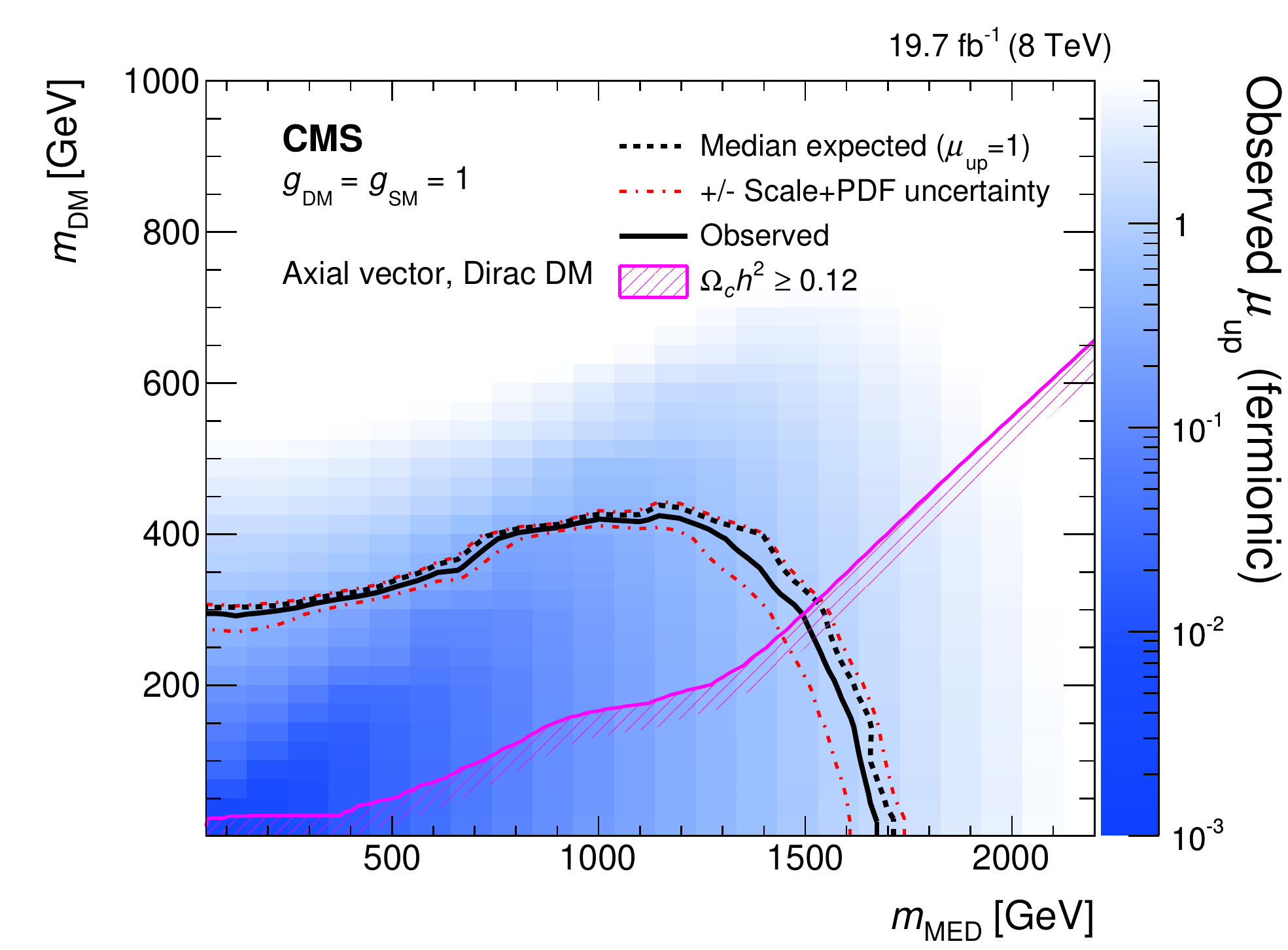}
\includegraphics[width=0.49\textwidth]{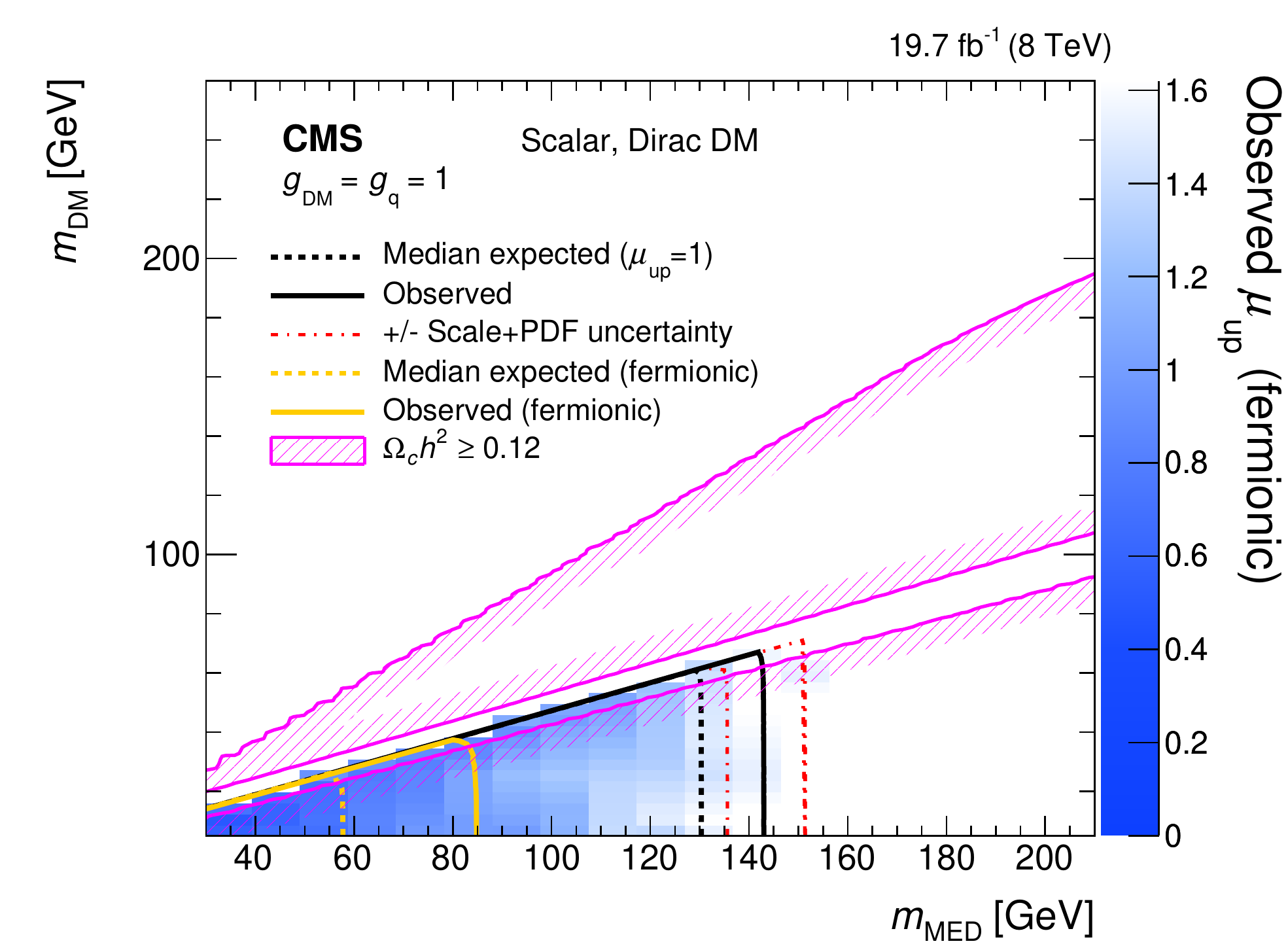}
\includegraphics[width=0.49\textwidth]{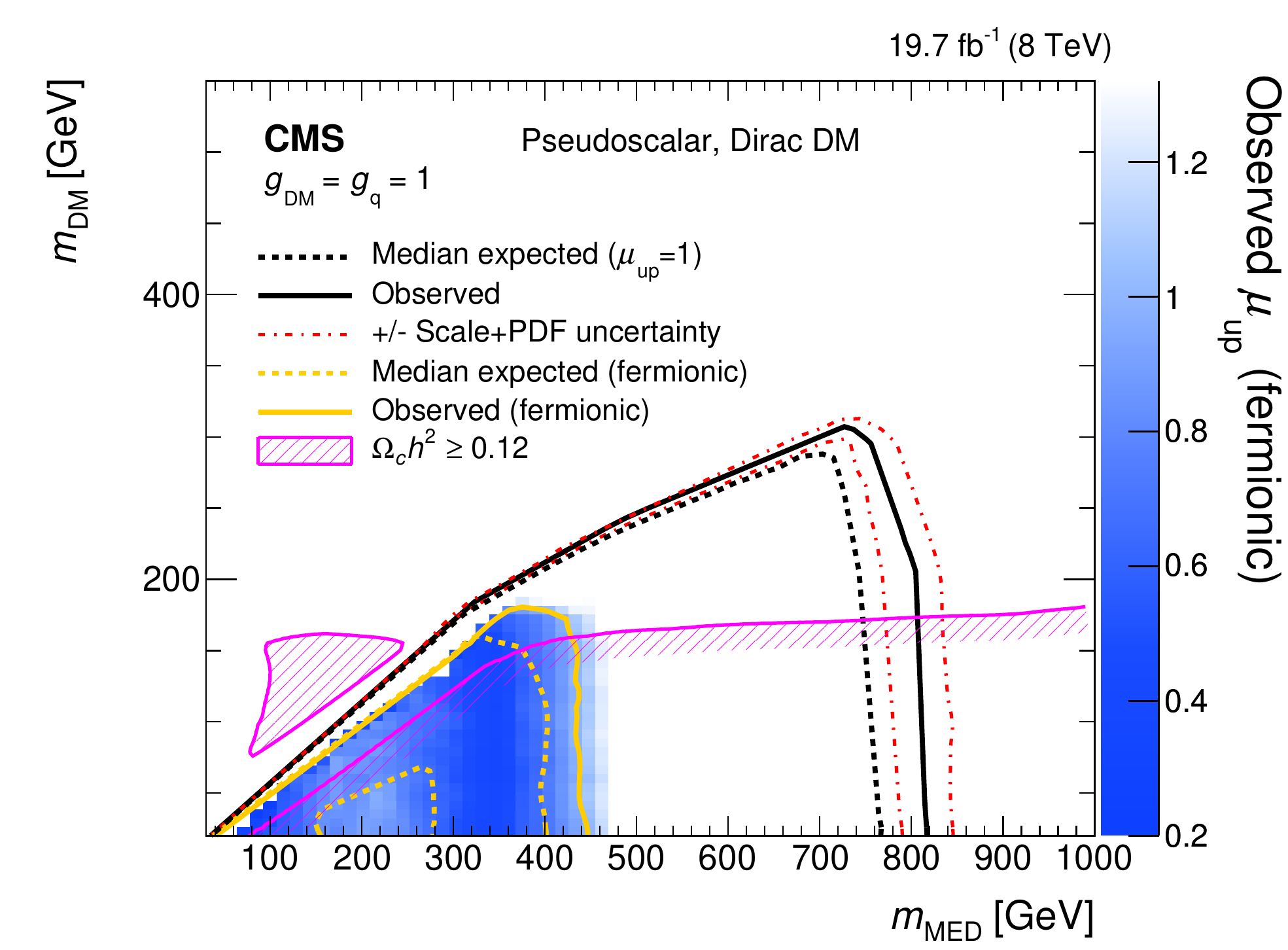}
\caption{The 90\% CL exclusion contours in the
$m_{\mathrm{MED}}-m_{\mathrm{DM}}$ plane assuming vector (top--left), axial vector (top--right),
scalar (bottom--left), and pseudoscalar (bottom--right) mediators.  The scale shown on the right hand axis
shows the expected
90\% CL exclusion upper limit on the signal strength, assuming the mediator only couples to
fermions.
For the scalar and pseudoscalar mediators, the exclusion contour
assuming coupling only to fermions (fermionic) is also shown. The
white region shows model points that are not tested when assuming coupling
only to fermions and are not expected to be excluded by this analysis under this
assumption. The red dot-dashed lines indicate the variation in the exclusion contours due to modifying
the renormalization and factorization scales by a factor of two in the generation of the signal.
In all cases, the excluded region is to the bottom--left of the contours,
except for the relic density, which shows the regions for which $\Omega_c h^2\ge0.12$, as indicated by the shading.
In all of the models, the mediator width is determined using the minimum width assumption\label{fig:masslims}.}
\end{figure}

Figure~\ref{fig:masslims} shows the 90\% CL exclusions for the vector,
axial vector, scalar, and pseudoscalar mediator models. The 90\%
upper limit on $\mu$ ($\mu_{\text{up}}$), when assuming that the mediator couples only to
fermions (fermionic), is shown by the blue color scale.
As described in Section~\ref{sec:signals},
the limits are calculated assuming a minimum width for the signal~\cite{An:2012va,Abercrombie:2015wmb,Fox:2011pm,simplified1}.
For the pseudoscalar interpretation, there is a region of
masses between 150 and 280\GeV for which the decrease in
cross section with larger mediator mass is balanced by an increase in acceptance for the
signal, so that the expected signal contribution remains roughly constant. The expected
value of $\mu_{\mathrm{up}}$ is larger than 1 in this region, resulting in an
``island'' at small $m_{\mathrm{DM}}$, where no exclusion is expected at the 90\% CL.
However, the observed value of $\mu_{\mathrm{up}}$ is smaller than 1 throughout this
region at 90\% CL, thus the island is not present in the observed limits.

The results are compared, for all four types of
mediators, to constraints obtained from the observed cosmological relic density
of DM as determined from measurements of the cosmic microwave background by the
WMAP and Planck experiments~\cite{Bennett:2003ba,Spergel:2006hy,Planck:2006aa}. The expected DM
abundance is estimated, separately for each model, using a thermal freeze-out
mechanism implemented in \textsc{MadDM}2.0.6~\cite{Backovic:2015cra}, and compared with the
observed cold DM density $\Omega_c h^2=0.12$~\cite{Ade:2013zuv}, as described in
Ref.~\cite{Pree:2016hwc}. It is assumed that the hypothesized simplified model
provides the only relevant dynamics for DM interaction beyond the SM.

\begin{figure}[htbp] \centering
\includegraphics[width=0.49\textwidth]{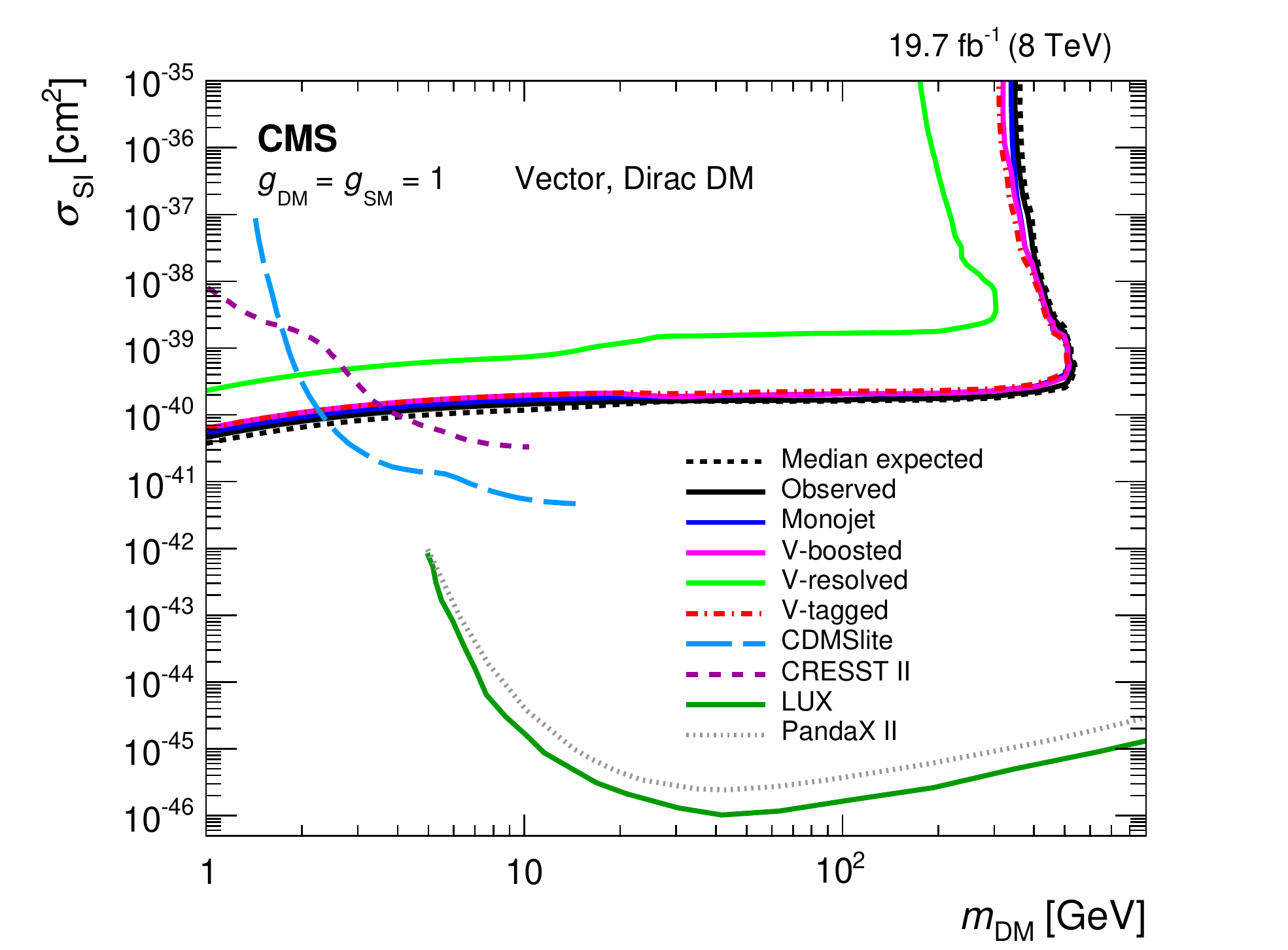}
\includegraphics[width=0.49\textwidth]{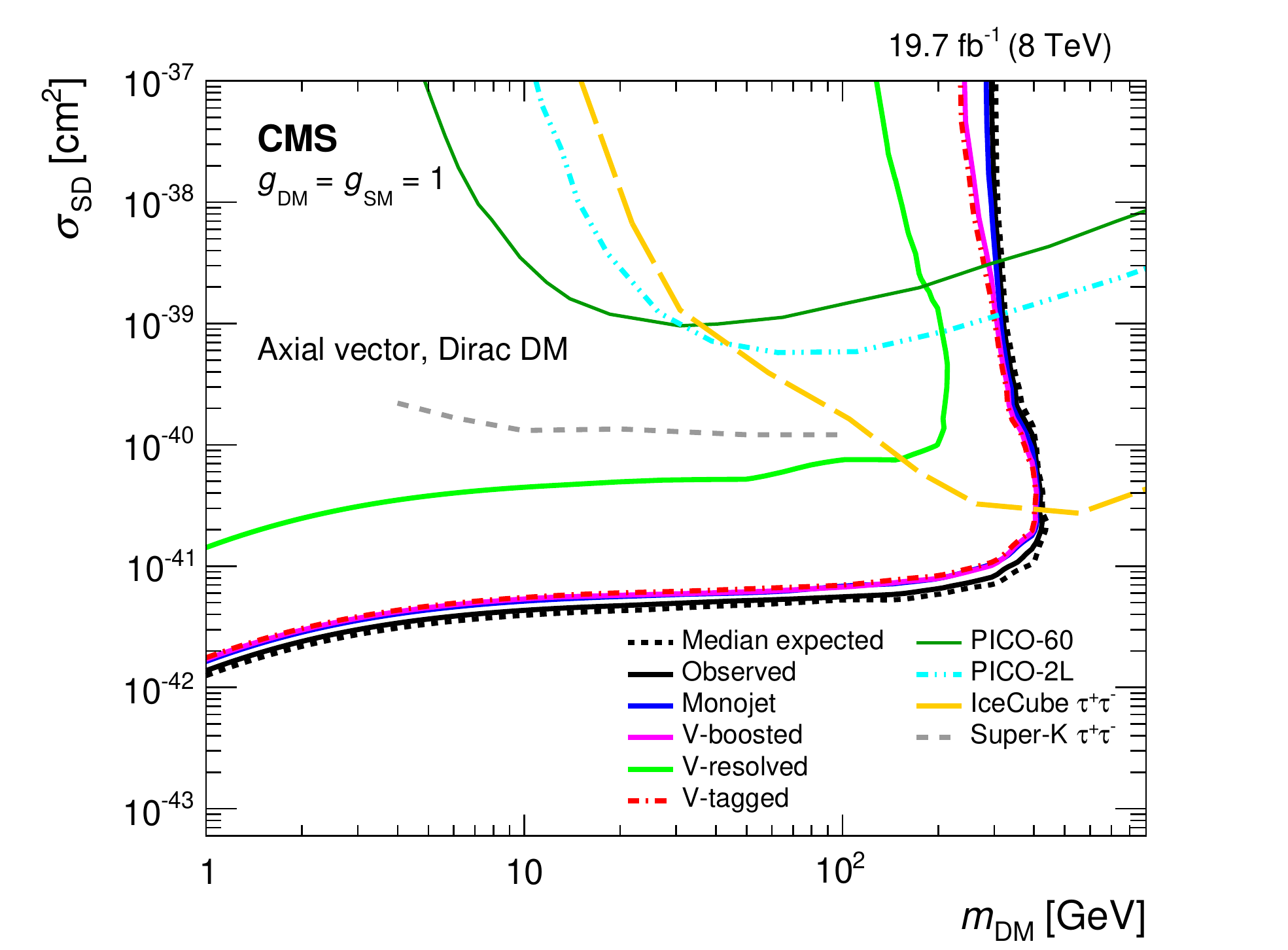}
\includegraphics[width=0.49\textwidth]{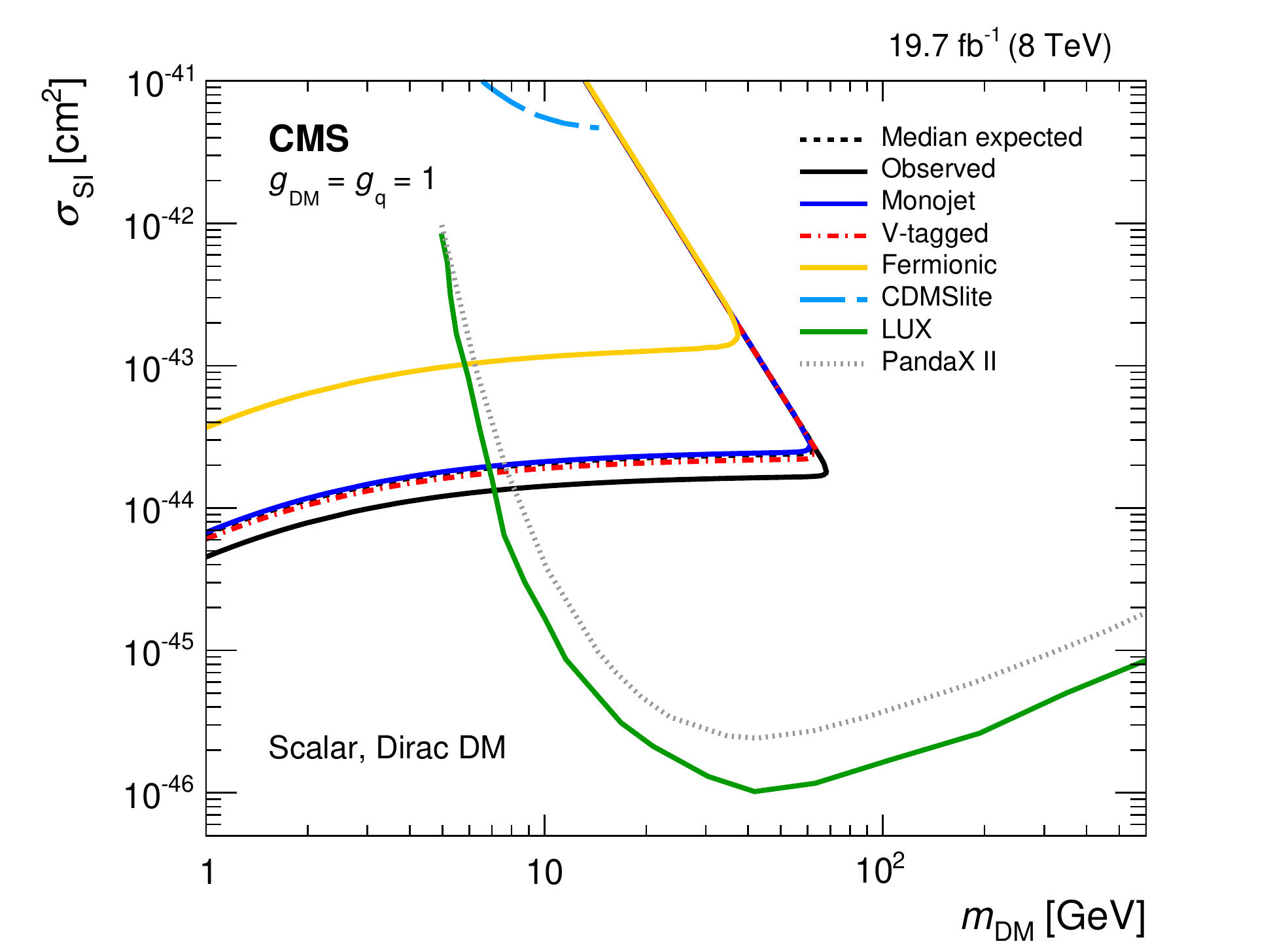}
\includegraphics[width=0.49\textwidth]{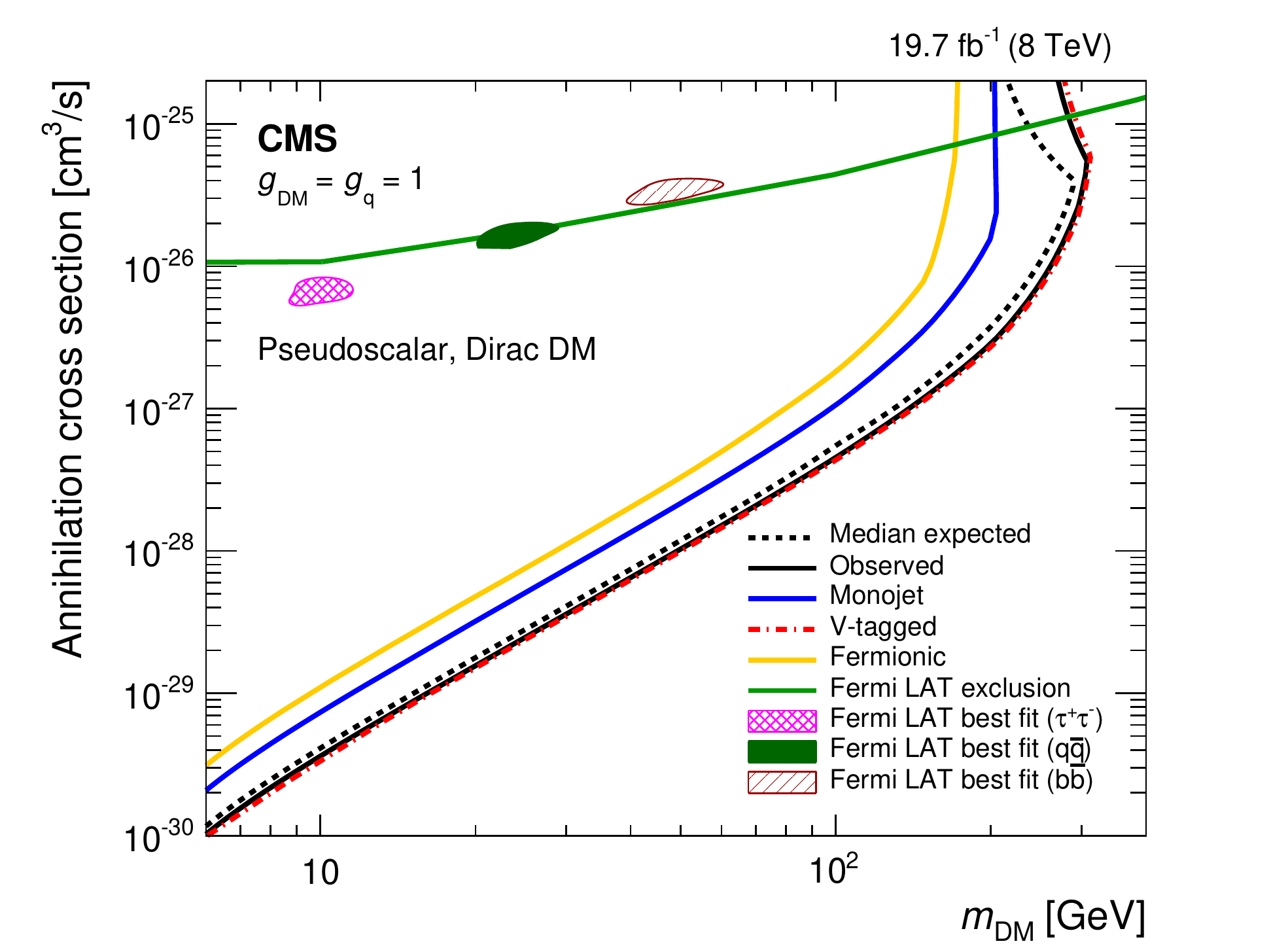}
\caption{The 90\% CL exclusion contours in the
$m_{\mathrm{DM}}-\sigma_{\mathrm{SI}}$ or $m_{\mathrm{DM}}-\sigma_{\mathrm{SD}}$
plane assuming vector (top--left), axial vector (top--right),
scalar (bottom--left) mediators. Also shown is the 90\% CL exclusion in DM annihilation cross
section as a function of $m_{\mathrm{DM}}$ for a pseudoscalar mediator (bottom--right).
For the scalar and pseudoscalar
mediators, the exclusion contours assuming the mediator only
couples to fermions (fermionic) is also shown.
The excluded region in all plots is to the top--left of the
contours for the results from this analysis while the DD experiments and Fermi LAT excluded regions
are above the lines shown. In the vector and axial vector models, limits are shown
independently for monojet, V-boosted, and V-resolved categories. The red dot-dashed line shows the partial
combination of the V-tagged categories for which the V-boosted category provides the dominant contribution.  In all
of the mediator models, a minimum mediator width is assumed. For the pseudoscalar mediator,
68\% CL preferred regions, obtained using data from Fermi LAT, for DM annihilation to
light-quarks ($\PQq\PAQq$), $\tau^{+}\tau^{-}$, and $\bbbar$ are given
by the solid green, hatched pink, and shaded brown coloured regions, respectively.\label{fig:xslims} }
\end{figure}

Figures~\ref{fig:xslims}(top--left),~\ref{fig:xslims}(top--right), and~\ref{fig:xslims}(bottom--left) show
the same exclusion contours, this time translated into the planes of
$m_{\mathrm{DM}}-\sigma_{\mathrm{SI}}$ or $m_{\mathrm{DM}}-\sigma_{\mathrm{SD}}$,
where $\sigma_{\mathrm{SI}}$ and $\sigma_{\mathrm{SD}}$ are
the SI or SD DM-nucleon scattering cross sections. These representations allow a more direct
comparison with limits from the DD experiments.
The translations are obtained following the procedures outlined in Ref.~\cite{Malik:2014ggr}
for the vector and axial vector mediators and in Refs.~\cite{Harris:2015kda,Boveia:2016mrp}
for the scalar mediator.
It should be noted that the limits set from this analysis are only valid
for the simplified model, and in particular that they assume
$g_{\mathrm{DM}}=g_{\mathrm{SM}}=g_{\PQq}=1$. For the scalar mediator model, it is assumed that
only heavy quarks (top and bottom) contribute. Such a choice limits the
sensitivity for DD experiments, however, it allows the direct comparison
between collider and DD experiments without an additional assumption for the
light-quark couplings~\cite{Harris:2015kda}. For the vector and scalar models, the limits are compared with those from the
LUX~\cite{Akerib:2016vxi}, CDMS lite~\cite{PhysRevLett.116.071301}, CRESST II~\cite{Angloher:2015ewa}, and PandaX II~\cite{Tan:2016zwf} experiments.
The limits from the LUX experiment currently
provides the strongest constraints on $\sigma_{\mathrm{SI}}$ for $m_{\rm{DM}} \gtrsim 4$\GeV,
while for values of $m_{\rm{DM}}<2$\GeV the analysis in this paper provides more stringent constraints on the vector and scalar models
as shown in Figs.~\ref{fig:xslims}(top--left) and Fig.~\ref{fig:xslims}(bottom--left), respectively.
For axial vector couplings, the limits are compared with DM--proton scattering limits
from the PICO-2L~\cite{Amole:2015lsj}, PICO-60~\cite{Amole:2015pla}, IceCube~\cite{Aartsen:2016exj},
and Super-Kamiokande~\cite{Choi:2015ara} experiments. In this model, the limits obtained in this analysis
are superior for DM masses up to $300$\GeV.

Pseudoscalar-mediated DM-nucleon interactions are suppressed at large velocities.
The most appropriate
comparison is therefore to the most sensitive bounds on indirect detection
from the Fermi LAT collaboration~\cite{Ackermann:2011wa,Abdo:2010ex}.
These limits apply to a scenario in which DM annihilates in the centre of a galaxy,
producing a $\gamma$ ray signature.
The signature can be interpreted as DM
annihilation to b quark pairs, allowing direct comparison with limits from
this analysis~\cite{Buchmueller:2015eea,Buckley:2014fba,Harris:2014hga}.

Figure~\ref{fig:xslims}(bottom--right) shows the exclusion contours assuming pseudoscalar
mediation in the plane of DM pair annihilation cross section versus
$m_{\mathrm{DM}}$. It is assumed that only heavy quarks
contribute in the production of the mediator, while for the
interpretation of the Fermi LAT limits in the annihilation cross section, it is assumed
that the mediator decays only to b quark pairs. As with all of the simplified model interpretations,
the DM particle is assumed to be a Dirac fermion.
The results shown from Fermi LAT have been scaled by a factor of two compared
to Ref.~\cite{Ackermann:2011wa}, because of the assumption of a Majorana DM
fermion made by that analysis. The limits from this analysis improve on those from Fermi LAT for DM masses up to 150\GeV.

An excess in $\gamma$ ray emission, consistent with the annihilation of DM, at the galactic
centre has been reported in several studies using data from Fermi LAT~\cite{TheFermi-LAT:2015kwa,Gordon:2013vta,Abazajian:2012pn,Hooper:2010mq}.
Further studies of this excess suggest that DM
annihilation could be mediated by a light pseudoscalar  particle~\cite{Agrawal:2014oha,Calore:2014nla}.
The 68\% CL preferred regions in this plane assuming the annihilation of DM
pairs to light-quarks ($\PQq\PAQq$), $\tau^{+}\tau^{-}$, or $\bbbar$,
using data from Fermi LAT, are shown as solid colour regions in Fig.~\ref{fig:xslims}(bottom--right).
For the simplified model, and assuming that $g_{\mathrm{DM}}=g_{\PQq}=1$,
all of these regions are excluded by this analysis.

\section{Summary}
A search has been presented for an excess of events with at least one
energetic jet in association with large \ETm in a
data sample of proton-proton collisions at a centre-of-mass energy of 8\TeV. The
data correspond to an integrated luminosity of 19.7\fbinv collected with the CMS
detector at the LHC. Sensitivity to a potential mono-V signature is
achieved by the addition of two event categories that select hadronically decaying V-bosons
using novel jet substructure techniques.
This search is the first at CMS to use jet substructure
techniques to identify hadronically decaying vector bosons in both
Lorentz-boosted and resolved scenarios.
The sensitivity of the search has been increased compared to the previous CMS result by using
the full shape of the $\ETm$ distribution to discriminate signal from standard model
backgrounds and by using additional data control regions.
No significant deviation is observed in the \ETm distributions relative to the expectation from
standard model backgrounds. The results of the search
are interpreted under a set of simplified models that describe the production
of dark matter (DM) particle pairs via vector, axial vector, scalar, or pseudoscalar mediation.
Constraints are placed
on the parameter space of these models. The search was the first at CMS to be interpreted using the
simplified models for DM production. The search excludes DM production via vector or axial
vector mediation with mediator masses up to 1.5\TeV, within the simplified model assumptions. When
compared to direct detection experiments, the limits from this analysis provide the strongest constraints at
small DM masses in the vector model and for DM masses up to 300\GeV in the axial vector model.
For scalar and pseudoscalar
mediated DM production, this analysis excludes mediator masses up to 80 and
400\GeV, respectively. The results of this analysis provide the strongest constraints on DM
pair annihilation cross section via a pseudoscalar interaction for DM masses
up to 150\GeV compared to the latest indirect detection results from Fermi LAT.

\begin{acknowledgments}
\hyphenation{Bundes-ministerium Forschungs-gemeinschaft Forschungs-zentren} We congratulate our colleagues in the CERN accelerator departments for the excellent performance of the LHC and thank the technical and administrative staffs at CERN and at other CMS institutes for their contributions to the success of the CMS effort. In addition, we gratefully acknowledge the computing centres and personnel of the Worldwide LHC Computing Grid for delivering so effectively the computing infrastructure essential to our analyses. Finally, we acknowledge the enduring support for the construction and operation of the LHC and the CMS detector provided by the following funding agencies: the Austrian Federal Ministry of Science, Research and Economy and the Austrian Science Fund; the Belgian Fonds de la Recherche Scientifique, and Fonds voor Wetenschappelijk Onderzoek; the Brazilian Funding Agencies (CNPq, CAPES, FAPERJ, and FAPESP); the Bulgarian Ministry of Education and Science; CERN; the Chinese Academy of Sciences, Ministry of Science and Technology, and National Natural Science Foundation of China; the Colombian Funding Agency (COLCIENCIAS); the Croatian Ministry of Science, Education and Sport, and the Croatian Science Foundation; the Research Promotion Foundation, Cyprus; the Secretariat for Higher Education, Science, Technology and Innovation, Ecuador; the Ministry of Education and Research, Estonian Research Council via IUT23-4 and IUT23-6 and European Regional Development Fund, Estonia; the Academy of Finland, Finnish Ministry of Education and Culture, and Helsinki Institute of Physics; the Institut National de Physique Nucl\'eaire et de Physique des Particules~/~CNRS, and Commissariat \`a l'\'Energie Atomique et aux \'Energies Alternatives~/~CEA, France; the Bundesministerium f\"ur Bildung und Forschung, Deutsche Forschungsgemeinschaft, and Helmholtz-Gemeinschaft Deutscher Forschungszentren, Germany; the General Secretariat for Research and Technology, Greece; the National Scientific Research Foundation, and National Innovation Office, Hungary; the Department of Atomic Energy and the Department of Science and Technology, India; the Institute for Studies in Theoretical Physics and Mathematics, Iran; the Science Foundation, Ireland; the Istituto Nazionale di Fisica Nucleare, Italy; the Ministry of Science, ICT and Future Planning, and National Research Foundation (NRF), Republic of Korea; the Lithuanian Academy of Sciences; the Ministry of Education, and University of Malaya (Malaysia); the Mexican Funding Agencies (BUAP, CINVESTAV, CONACYT, LNS, SEP, and UASLP-FAI); the Ministry of Business, Innovation and Employment, New Zealand; the Pakistan Atomic Energy Commission; the Ministry of Science and Higher Education and the National Science Centre, Poland; the Funda\c{c}\~ao para a Ci\^encia e a Tecnologia, Portugal; JINR, Dubna; the Ministry of Education and Science of the Russian Federation, the Federal Agency of Atomic Energy of the Russian Federation, Russian Academy of Sciences, and the Russian Foundation for Basic Research; the Ministry of Education, Science and Technological Development of Serbia; the Secretar\'{\i}a de Estado de Investigaci\'on, Desarrollo e Innovaci\'on and Programa Consolider-Ingenio 2010, Spain; the Swiss Funding Agencies (ETH Board, ETH Zurich, PSI, SNF, UniZH, Canton Zurich, and SER); the Ministry of Science and Technology, Taipei; the Thailand Center of Excellence in Physics, the Institute for the Promotion of Teaching Science and Technology of Thailand, Special Task Force for Activating Research and the National Science and Technology Development Agency of Thailand; the Scientific and Technical Research Council of Turkey, and Turkish Atomic Energy Authority; the National Academy of Sciences of Ukraine, and State Fund for Fundamental Researches, Ukraine; the Science and Technology Facilities Council, UK; the US Department of Energy, and the US National Science Foundation.

Individuals have received support from the Marie-Curie programme and the European Research Council and EPLANET (European Union); the Leventis Foundation; the A. P. Sloan Foundation; the Alexander von Humboldt Foundation; the Belgian Federal Science Policy Office; the Fonds pour la Formation \`a la Recherche dans l'Industrie et dans l'Agriculture (FRIA-Belgium); the Agentschap voor Innovatie door Wetenschap en Technologie (IWT-Belgium); the Ministry of Education, Youth and Sports (MEYS) of the Czech Republic; the Council of Science and Industrial Research, India; the HOMING PLUS programme of the Foundation for Polish Science, cofinanced from European Union, Regional Development Fund, the Mobility Plus programme of the Ministry of Science and Higher Education, the National Science Center (Poland), contracts Harmonia 2014/14/M/ST2/00428, Opus 2013/11/B/ST2/04202, 2014/13/B/ST2/02543 and 2014/15/B/ST2/03998, Sonata-bis 2012/07/E/ST2/01406; the Thalis and Aristeia programmes cofinanced by EU-ESF and the Greek NSRF; the National Priorities Research Program by Qatar National Research Fund; the Programa Clar\'in-COFUND del Principado de Asturias; the Rachadapisek Sompot Fund for Postdoctoral Fellowship, Chulalongkorn University and the Chulalongkorn Academic into Its 2nd Century Project Advancement Project (Thailand); and the Welch Foundation, contract C-1845.
\end{acknowledgments}

\bibliography{auto_generated}

\appendix
\section{Supplementary Material\label{app:suppMat}}
\begin{figure}[hbtp]
\centering
{\includegraphics[width=\textwidth]{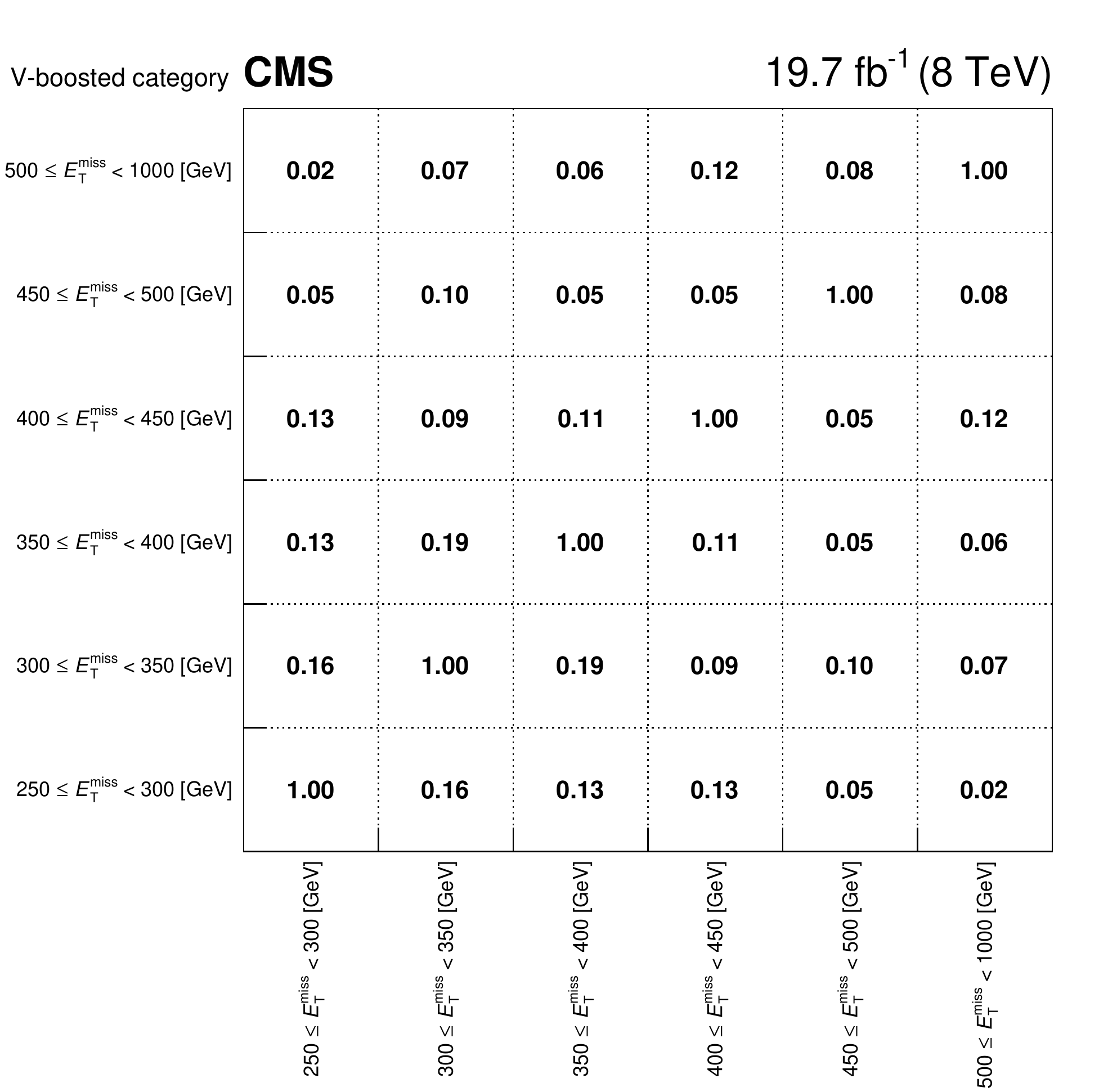}}
\caption{
Correlations between the predicted number of background events in each bin of \ETmiss in the V-boosted category. 
The correlation is determined from the simultaneous fit to data in the dimuon, single muon, and photon control regions 
in all the three event categories. 
\label{fig:correlation_VB} }
\end{figure}

\begin{figure}[hbtp]
\centering
{\includegraphics[width=\textwidth]{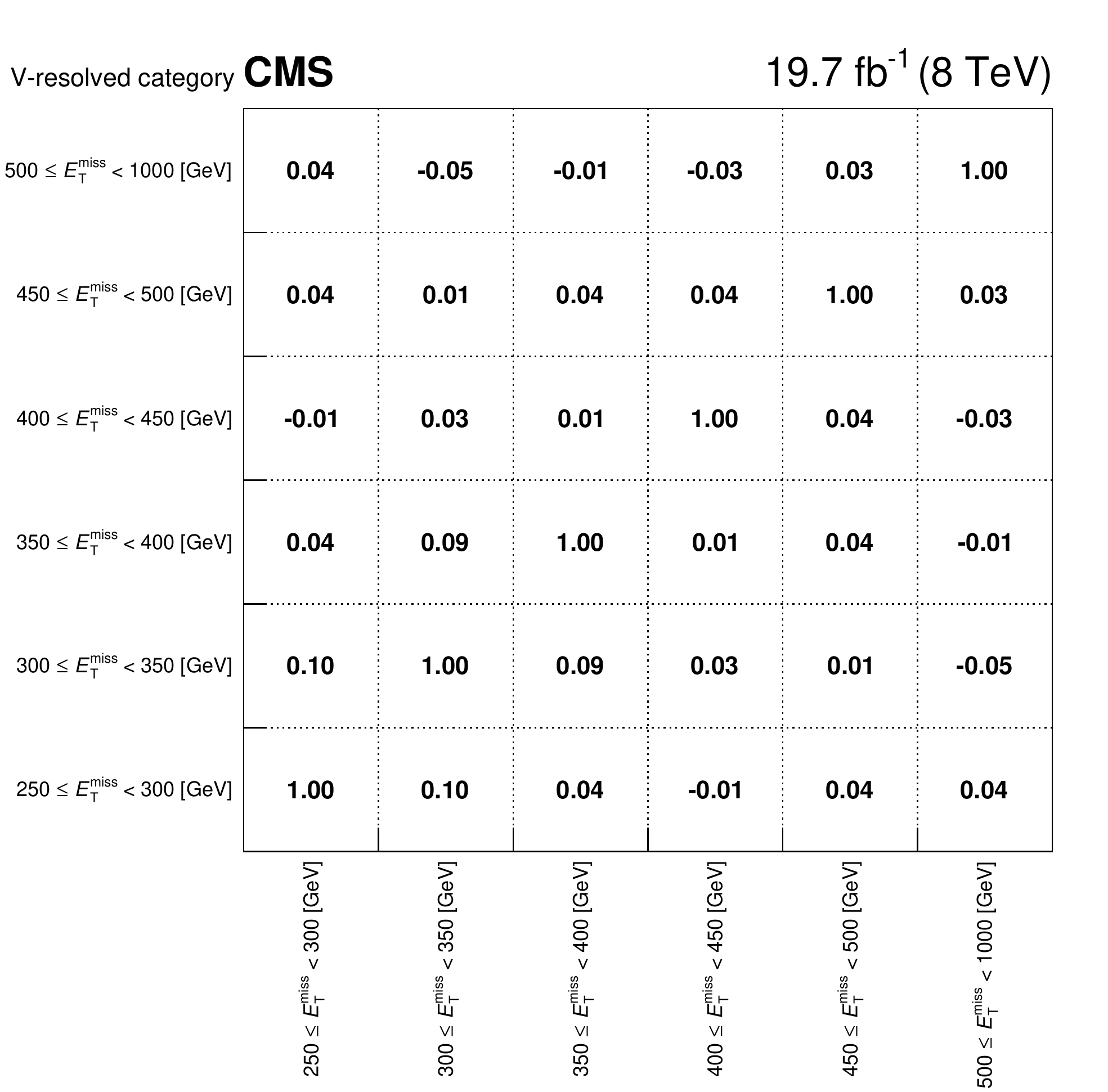}}
\caption{
Correlations between the predicted number of background events in each bin of \ETmiss in the V-resolved category. 
The correlation is determined from the simultaneous fit to data in the dimuon, single muon, and photon control regions 
in all the three event categories. 
\label{fig:correlation_VR} }
\end{figure}

\begin{figure}[hbtp]
\centering
{\includegraphics[width=\textwidth]{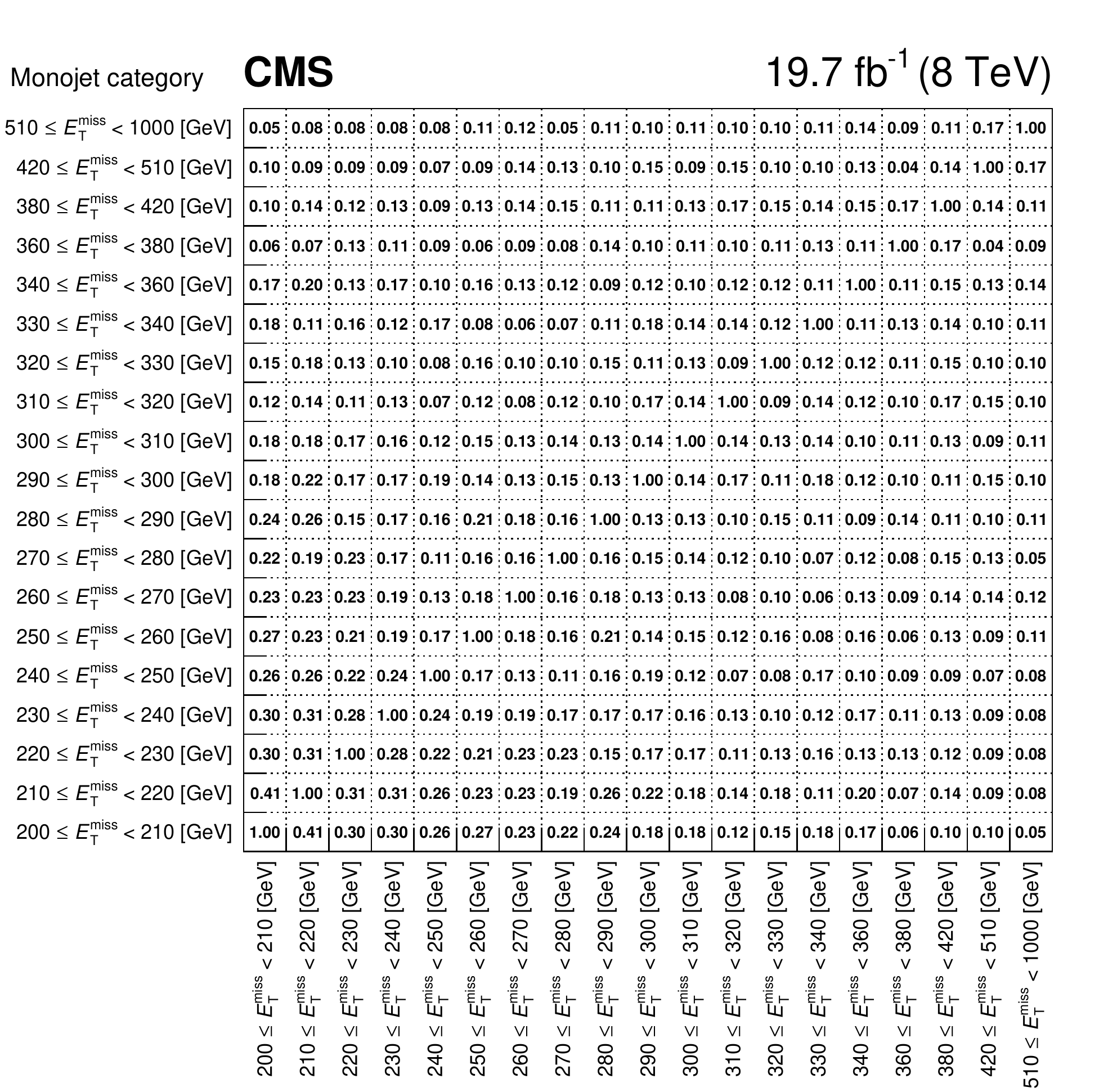}}
\caption{
Correlations between the predicted number of background events in each bin of \ETmiss in the monojet category. 
The correlation is determined from the simultaneous fit to data in the dimuon, single muon, and photon control regions 
in all the three event categories. 
\label{fig:correlation_MJ} }
\end{figure}

\cleardoublepage \section{The CMS Collaboration \label{app:collab}}\begin{sloppypar}\hyphenpenalty=5000\widowpenalty=500\clubpenalty=5000\textbf{Yerevan Physics Institute,  Yerevan,  Armenia}\\*[0pt]
V.~Khachatryan, A.M.~Sirunyan, A.~Tumasyan
\vskip\cmsinstskip
\textbf{Institut f\"{u}r Hochenergiephysik der OeAW,  Wien,  Austria}\\*[0pt]
W.~Adam, E.~Asilar, T.~Bergauer, J.~Brandstetter, E.~Brondolin, M.~Dragicevic, J.~Er\"{o}, M.~Flechl, M.~Friedl, R.~Fr\"{u}hwirth\cmsAuthorMark{1}, V.M.~Ghete, C.~Hartl, N.~H\"{o}rmann, J.~Hrubec, M.~Jeitler\cmsAuthorMark{1}, A.~K\"{o}nig, I.~Kr\"{a}tschmer, D.~Liko, T.~Matsushita, I.~Mikulec, D.~Rabady, N.~Rad, B.~Rahbaran, H.~Rohringer, J.~Schieck\cmsAuthorMark{1}, J.~Strauss, W.~Treberer-Treberspurg, W.~Waltenberger, C.-E.~Wulz\cmsAuthorMark{1}
% ...
\vskip\cmsinstskip
% ...
1:~~Also at Vienna University of Technology, Vienna, Austria\\
% ...

\end{sloppypar}
\end{document}